\documentclass{iopart}
\usepackage{iopams}  
\usepackage{amssymb}
\usepackage{braket}
\usepackage{mathrsfs}
\usepackage{graphicx}
\usepackage[colorlinks=true,citecolor=blue,linkcolor=black]{hyperref}
\usepackage{xcolor}
\usepackage{caption}
\usepackage{subcaption}
\usepackage{tikz}
\usetikzlibrary{shapes}
\usepackage{array}
\newcolumntype{L}{>{$}l<{$}} 

\newcommand{\red}[1]{\textcolor{black}{#1}}

\newcommand{\blue}[1]{\textcolor{black}{#1}}

\begin{document}
\title{Multi-mode architectures for noise-resilient superconducting qubits} 
\author{Alessio Calzona$^1$ and Matteo Carrega$^{2}$}
\address{$^1$ Institute for Theoretical Physics and Astrophysics, University of W\"urzburg, 97074 W\"urzburg, Germany}
\address{$^2$ CNR-SPIN, Via Dodecaneso 33, 16146 Genova, Italy}
\begin{abstract}
Great interest revolves around the development of new strategies to efficiently store and manipulate quantum information in a robust and decoherence-free fashion.
Several proposals have been put forward to encode information into qubits that are simultaneously insensitive to relaxation and to dephasing processes. Among all, given their versatility and high-degree of control, superconducting qubits have been largely investigated in this direction. Here, we present a survey on the basic concepts and ideas behind the implementation of novel superconducting circuits with intrinsic protection against decoherence at a hardware level. In particular, the main focus is on multi-mode superconducting circuits, the paradigmatic example being the so-called $0-\pi$ circuit. We report on their working principle and possible physical implementations based on conventional Josephson elements, presenting recent experimental realizations, discussing both fabrication methods and characterizations.
\end{abstract}
\section{Introduction}
\label{intro}
In recent years tremendous progress has been made in the realization and manipulation of superconducting circuits for quantum technology applications~\cite{acin2018, devoret2013, krantz2019}. Due to the steady increase in coherence times, precise control, and potential scalability~\cite{devoret2013, krantz2019, wendin2017, girvin2009}, these superconducting systems are among the best candidate for the realization of quantum computers~\red{\cite{acin2018, nielsen_book, benenti_book, you2005, moore_ibm2017, rigetti2018, vozhakov2022}}. Superconducting qubits are built from macroscopic-sized superconducting circuits that behave as quantum anharmonic oscillators. Indeed, their three elementary building blocks are linear capacitors, inductors, and Josephson junctions (JJs)~\cite{devoret2013, josephson1962}, the last being responsible for anharmonicity in the spectrum. Importantly, the circuit parameters, which in turn determine the spectral properties of the superconducting qubits, are easily tunable and offer a rich variety of useful operating regimes. Several circuit designs have been envisioned and realized over the last two decades~\cite{devoret2013, wendin2017, devoret2004, clarke2008, mooij1999, oliver2013, koch2007, manucharyan2009, casparis2018}, each one with its own strengths and limitations.\\

One of the main challenges associated with the realization of quantum computers is to cope with the detrimental effects of decoherence. Indeed, due to the unavoidable coupling with the surrounding environment, qubits are subject to several sources of noise~\cite{krantz2019, weiss_book, schnirman2002, vion2002, burkard2004, faoro2006, clerk2010, carrega2020}, which introduce errors in quantum information processing. In the current NISQ (Noisy Intermediate Scale quantum) era \cite{preskill2018, arute2019, hsu_ces_2018}, performances of quantum processors are strongly limited by decoherence. In particular, only shallow codes, i.e. with a small number of time steps, can be safely run to avoid quantum information being completely overwhelmed by noise~\cite{preskill2018}. To overcome this issue, error correcting schemes, able to actively detect faulty qubit behaviors and restore the right quantum state, have been proposed~\cite{knill2001, aharonov2008, fowler2012, kelly2015, barends2014, gambetta2017, kapit2018, wootton2020, magesan2012, temme2017}. The actual implementation of these schemes, which would mark the beginning of the new era of fault-tolerant quantum computers, is however quite challenging. To be effective, it indeed requires large hardware overhead~\cite{fowler2012, devitt2013, ogorman2017} and, crucially, systems that already feature very low noise levels. Therefore, in the last years, huge theoretical and experimental efforts have been put into the search for noise protection at a hardware level. For instance, \red{ early attempts to create protected qubits were pursued exploiting Josephson contacts with non-trivial pairing and current-phase relation~\cite{ioffe99, amin2005, klenov2006}. Moreover,} the great interest in emergent topological states of matter exactly stems from their potential in building fault-tolerant protected qubits~\cite{haldane2017, stern2013, nayak2008, carrega2021, alicea2016}.\\

Several routes for noise mitigation of superconducting qubits are currently under inspection, both at a software and hardware level. Regarding the former, in addition to the development of active error correction schemes, recent years have witnessed the development of quantum error mitigation (QEM) strategies, \blue{which include, for instance,} a collection of techniques that can reduce the impact
of errors at the cost of performing extra measurements and data post-processing~\cite{temme2017, li2017, endo2018, suzuki2022, endo2021, koczor2021, bultrini2021, calzona2022}, already successfully implemented in different experiments~\cite{kandala2019, saga2019, arute2020}. \blue{Other approaches include the exploitation of dynamical decoupling sequences \cite{paladino2014, bylander2011,uhrig2007,uys2009} as well as the recent implementation of the so-called dynamical sweet spots, which has been successfully used to mitigate errors in both superconducting~\cite{didier2022} and spin~\cite{bertaina2020} qubits.}
At a hardware level, two complementary approaches are currently pursued. One is the direct minimization of the noise amplitude, usually achieved by improved fabrication techniques and optimization of material engineering~\cite{krantz2019, kandala2017, neill2018}. The second aims at improving the qubit design, by relying on novel geometries, in order to reduce the sensitivity both to charge and flux noise~\cite{doucot2012, gyenis2021, dannon2022, ioffe2002, doucot2002, kitaev2006, brooks2013, smith2020, gladchenko2009, bell2014, pechenezhskiy2020, kalashnikov2020, gyenis2021_exp, chirolli2021}. This review mainly focuses on this last approach. Specifically, we will discuss novel circuit architectures featuring peculiar computational Hilbert spaces which makes them potentially insensitive to noise. We will focus on the so-called multi-mode circuits, which consist of new device designs with a larger number of degrees of freedom with respect to conventional qubits such as transmon~\cite{koch2007} or fluxonium ones~\cite{manucharyan2009} (which are usually described as single-mode circuits). There, logical wavefunctions are encoded in different subspaces with non-local nature, which allow for simultaneous exponential suppression of both charge and flux noise.\\

Multi-mode based qubit architectures  can be described within the standard framework widely used for conventional superconducting qubits. This is the so-called circuit quantum electrodynamics (QED), an active field that evolved from cavity QED \cite{kimble1998,haroche2006, blais2020} and that explores the interaction between light and (artificial) matter.  Here, we do not aim at giving an exhaustive discussion of the vast theoretical and experimental panorama in this field. Excellent reviews covering different aspects of circuit QED and conventional superconducting qubits can be found in the existing literature, for example, see Refs.~\cite{wendin2017, girvin2009, clarke2008, blais2004, gu2017,vool2017, blais2021}. It is worth underlining that standard superconducting circuits, such as transmon and fluxonium, are well described by single-mode circuit in this framework. As mentioned above, in this work we will introduce and discuss novel qubit designs that require a multi-mode description, and we will present several experimental realizations that have demonstrated their potential and protection against noise sources. It is important to mention that such architectures beyond the single-mode qubits are still less mature with respect to state-of-the-art qubits, such as transmon~\cite{koch2007, roth2022} or fluxonium~\cite{ manucharyan2009, somoroff2021}, and come with non-trivial experimental challenges. Nevertheless, impressive experimental progress has been made in the last few years, demonstrating the huge potential of these new technologies. Further experimental breakthroughs in terms of fabrication and optimization could eventually unleash the full potential of multi-mode qubits, which can therefore become \red{important elements, complementary to single-mode circuits,} for the development of new quantum processors.\\  

The rest of this review is organized as follows. In \Sref{sec:sup_quantum_circuits} we introduce the basic concepts and general framework, discuss the main mechanisms that lead to dissipation and decoherence, and present the state-of-the-art single-mode superconducting circuits. In \Sref{sec:multimode} an extension to multi-mode circuits is given, with a particular focus on the paradigmatic example of a $0-\pi$ circuit~\cite{brooks2013}. \Sref{sec:experiments} is devoted to the discussion of recent experimental realizations of superconducting multi-mode circuits and the characterization of their performance. Finally, \Sref{sec:concl} is devoted to conclusions and perspectives.
\section{Basics of superconducting qubits}
In this Section we review some basic aspects, introducing useful quantities and how to describe a superconducting quantum circuit as a qubit. After discussing the concept of qubit decoherence,  we comment on the performance of the most investigated state-of-the-art single-mode qubits, i.e. transmon and fluxonium qubits. This preparatory Section paves the way for the investigation of multi-mode qubits design, both at the theoretical and experimental level, that are the subject of the next Sections.

\subsection{Superconducting quantum circuits}
\label{sec:sup_quantum_circuits}
Despite being macroscopic objects consisting of an extremely large number of atoms, superconducting quantum circuits feature the low-energy physics which resembles the one of simple few-body particles. This remarkable fact stems from few important points. The use of superconducting electrodes at very low temperature, typically a few tens of mK, allows for dissipationless transport and forces all the electrons to condense into Cooper pairs. This, together with the fact that charge density fluctuations are characterized by high energy scales (characteristic plasma frequencies are in the optical range~\cite{krantz2019, barone_book}), freezes almost all the degrees of freedom of the in principle many-body system, leaving with a handful of variables required to describe the low-energy dynamics of the system. Moreover, the description of superconducting circuits can be further simplified by the fact that their physical size does not exceed tens of $\mu$m, and it is much smaller than the wavelength corresponding to their typical frequency range, which is of the order of few GHz~\cite{devoret2013, girvin2009, blais2021}.
It is therefore possible to model a superconducting quantum circuit as a network of lumped two-terminal elements~\cite{girvin2009,vool2017} of capacitive or inductive nature. An optical micrograph image of a representative example of a superconducting circuit is reported in \Fref{fig:circuits}(a), while lumped-element schematics of some relevant qubit designs are reported in the remaining panels.

\subsubsection{Circuital elements}
Each basic circuital element $b$ (also known as a branch) can be characterized by the dissipationless current $i_b$ flowing through it and the voltage drop across it $v_b$. In view of an effective Hamiltonian description of the system, it is convenient to introduce the so-called branch fluxes $\Phi_b(t)$, defined as the time integral of the relative branch voltages
\begin{equation}
	\Phi_b(t) = \int_{-\infty}^{t} v_b(\tau) d\tau. 
\end{equation}
	This provides the key ingredients to analyze linear circuit elements. The energy stored in a linear capacitor with capacity $C_b$ depends quadratically on the voltage drop and it reads
\begin{equation}
	\label{eq:hc}
	T_b = \frac{1}{2} {C_b} v_b^2 = \frac{1}{2} {C_b} \dot\Phi_b^2.
\end{equation}
Similarly, the energy stored in a linear inductor is quadratic in the branch flux:
\begin{equation}
	\label{eq:hlL}
	U_b^l= \frac{\Phi_b^2}{2L_b}  ,
\end{equation}
where $L_b$ is the inductance which relates the branch current to the branch flux as $i_b = \Phi_b/{L_b}$.\\ 

The realization of qubits, however, requires the presence of non-linear elements to provide necessary anharmonicity in the energy spectrum. It is indeed crucial that the transition frequency between the two chosen energy levels that store quantum information largely differs from all the other excitation energies. This guarantees the possibility to properly address the computational space, safely neglecting other higher energy levels. In superconducting circuits, non-linear elements are naturally provided by JJs~\cite{josephson1962, barone_book}, which can be modeled as a pure Josephson element in parallel with a linear capacitor. A Josephson element describes the tunneling of Cooper pairs between two superconducting electrodes with superconducting phase difference $\chi$. It is characterized by the two Josephson equations
\begin{eqnarray}
	i_b(t) &= E_J \frac{2\pi}{\Phi_0} \sin(\chi)\\
	\frac{\partial \chi}{\partial t} &= \frac{2\pi}{\Phi_0} v_b(t),
\end{eqnarray}
where $\Phi_0 = h/(2e)$ is the superconducting flux quantum and $E_J$ is known as the Josephson energy. The second Josephson equation immediately relates the superconducting phase difference with the branch flux,
$\chi = 2\pi \Phi_b/\Phi_0$ (mod $2\pi$), while the first one represents the non-linear constitutive relation between supercurrent and the branch flux. The total energy stored in a Josephson element can be therefore evaluated as
\begin{equation}	
	\label{eq:hl}
	U_b^J = \int_{-\infty}^{t} i_b(\tau) \dot \Phi_b(\tau) \, d\tau = E_J \left[
	1- \cos(2\pi \Phi_b/\Phi_0)
	\right].
\end{equation}
Notice that the inductive nature of Josephson elements has a kinetic origin rather than a geometric one, as it directly stems from the kinetic energy of the Cooper pairs that tunnel between the superconducting electrodes. As we will see in \Sref{sec:superinductor}, the kinetic inductance of JJs is also one of the key ingredients for the realization of ``superinductors'', i.e. novel inductive circuital elements with very high impedance able to exceed the resistance quantum at the relevant ({microwave}) frequency, and low dissipation~\cite{manucharyan2012}.\\ 
\begin{figure}
	\centering
	\includegraphics[width=.7\textwidth]{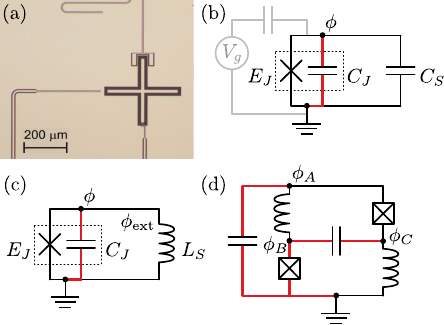}
	\caption{(a) Optical micrograph of a superconducting transmon qubit, showing the superconducting Al layer (light) and the exposed sappire substrate (dark). The large cross-shaped element, typical of the Xmon design, is the capacitor that shunts the (tunable) JJ, localized at the end of the bottom arm. Adapted from Ref.~\cite{barends2013}. (b) Electrical circuit of a capacitively shunted JJ. An external control line to change the offset charge of the floating islands is depicted in gray. The spanning \red{tree} connecting the single active node with the reference one is highlighted in red. (c) Electrical circuit of an inductively shunted JJ, with the spanning tree highlighted in red. (d) Example of the electrical circuit of a multi-mode device, the so-called $0-\pi$ qubit.
 Each crossed square represents a JJ, consisting of a Josephson element and a self-capacitance. Differently from the circuits in (b) and (c), it features three different active nodes, named $\phi_A$, $\phi_B$, and $\phi_C$, connected to the reference one via the spanning tree (in red).}
	\label{fig:circuits}
\end{figure}

In general, the branch fluxes $\Phi_b$ just introduced are not completely independent from each other because of the constraints imposed by the circuit topology. The latter is expressed by the use of Kirchhoff laws. A convenient way to properly define a set of independent degrees of freedom is based on the analysis of circuit nodes. We now highlight its key ideas while more detailed discussion can be found for example in Ref. \cite{devoret2013,vool2017,blais2021}. First of all, one has to define a spanning tree, i.e. select a set of branches connecting all nodes without forming any loop. The choice of the spanning tree is not unique (it is akin to the choice of a specific electromagnetic gauge) but it is generally convenient to choose a tree consisting only of capacitive elements. The existence of these particular types of trees is guaranteed in realistic circuits by the unavoidable presence of stray capacitances~\cite{vool2017,peikari1974}. Some examples of spanning trees are shown with thicker red lines in \Fref{fig:circuits}(b,c,d). The branches not included in the spanning tree are known as closure branches and each one of them defines a unique loop, obtained while adding the closure branch to the spanning tree.
Once a reference node is (arbitrarily) picked, it is possible to define for all the remaining ones a node voltage $v_n(t)$, which is measured with respect to the reference node, along the spanning tree. The time integrals of these node voltages are known as node fluxes
\begin{equation}
	\Phi_n(t) = \int_{-\infty}^{t} v_n(\tau) d\tau 
\end{equation}
and they represent the degrees of freedom of the superconducting circuit\footnote{A more detailed analysis should distinguish between active and passive nodes, the latter being the ones where only capacitors meet. Since the charge of the passive nodes is conserved, they have no dynamics and do not count as dynamical degrees of freedom. Moreover, note that the present approach is well defined and holds only as long as all the capacitors are linear: the study of advanced circuits with both capacitive and inductive non-linear elements is the subject of ongoing research~\cite{vool2017}.}. Their knowledge allows indeed to reconstruct the branch fluxes associated with every element of the circuit. In particular, if a branch belongs to the spanning tree and connects nodes $i$ and $j$, the branch flux is simply given by the difference $\Phi_b = \Phi_i-\Phi_j$. As for branch closures, in computing their flux one should also take into account the external flux $\Phi_{\rm ext}$ that might pierce the associated loop, leading to $\Phi_b = \Phi_i-\Phi_j+\Phi_{\rm ext}$. In \Fref{fig:circuits} (b-c), two representative circuits are characterized by a single node flux and they thus feature a single degree of freedom. By contrast, the more complex circuit depicted in \Fref{fig:circuits} (d) has three degrees of freedom and it represents a prototypical example of multi-mode qubits that we will discuss in the next Section.\\ 

With these definitions in hand, we can now sketch the procedure that leads to a quantum description of the circuit. More in-depth analysis can be found in excellent reviews such as Ref.~\cite{devoret2013,girvin2009, vool2017}. Looking at the form of the energies in Eq.~\eref{eq:hc}, \eref{eq:hlL} and \eref{eq:hl}, one can readily draw a mechanical analogy, where the flux nodes play the role of position coordinates, the electrostatic energy plays the role of kinetic energy (with the capacity $C$ playing the role of a mass) and the magnetic energy plays the role of the potential energy. It is therefore straightforward to describe the whole system in terms of the Lagrangian, summing over all the inductive and capacitive branches, 
\begin{equation}
	\mathcal{L} = \sum T_b - \sum U_b.
\end{equation}
where the sum extends over all branches. 
This allows introducing the conjugate momenta of the node fluxes
\begin{equation}
	Q_n = \frac{\partial \mathcal{L}}{\partial \dot\Phi_n} = \sum_m C_m (\dot \Phi_n - \dot \Phi_m). 
\end{equation} 
These quantities are called node charges, as they correspond to the algebraic sum of all the charges on the capacitors that connect the node $n$ with the neighbor nodes $m$. One can then write the Hamiltonian of the system $H(\Phi_n, Q_n)$ via the usual Legendre transform 
\begin{equation}
	\label{eq:Legendre}
	H(\Phi_n, Q_n) = \sum_n Q_n \dot \Phi_n- \mathcal{L}(\Phi_n,\dot \Phi_n)
\end{equation}and quantize it by promoting the conjugated variables $(\Phi_n, Q_n)$ to non-commuting quantum operators satisfying
\begin{equation}
	[\Phi_n, Q_m] = i \hbar \delta_{nm}.
\end{equation}

\subsubsection{Single-mode superconducting qubits}
The simplest realizations of superconducting qubits are based on circuits with one single degree of freedom (i.e. one single active node) and are thus called ``single-mode" qubits. The two basic architectures used for the realization of single-mode superconducting qubits are the so-called capacitively-shunted Josephson junction (CSJJ), depicted in \Fref{fig:circuits}(b), and the so-called inductively-shunted Josephson junction (ISJJ), depicted in \Fref{fig:circuits}(c). Given their importance, as stand-alone circuits but also as building blocks for multi-mode architectures, we now highlight their main features.\\ 

Let us analyze, at first, the CSJJ architecture. As shown in \Fref{fig:circuits}(b), a JJ with Josephson energy $E_J$ and self-capacitance $C_J$ is additionally shunted by a capacitor {$C$}. Its straightforward quantization leads to the Hamiltonian of an anharmonic oscillator, akin to a pendulum in a gravitational field~\cite{girvin2009,vool2017}
\begin{equation}
	\label{eq:CPB}
	H_{\rm CS} = - E_J \cos(\phi) + 4E_C (n - n_g)^2.
\end{equation} 
Here, we have introduced the dimensionless flux $\phi = 2\pi \Phi/\Phi_0$ and charge $n = Q/(2e) = -i\partial_\phi$, which obeys the canonical commutation relation $[\phi,n]=i$. The anharmonicity is provided by the cosinusoidal potential energy of the Josephson element. The operator $n$ counts the number of Cooper pairs that tunnel into the floating superconductive island, whose charging energy is proportional to the inverse of the total capacitance $E_C = e^2/[2(C_J +C)]$. In \Eref{eq:CPB}, we have also added a real parameter $n_g$, which takes into account the effect of an external potential on the charging energy of the superconductive island. Such a potential can be associated, for example, with the presence of an external gate voltage, sketched in gray in \Fref{fig:circuits}(b), that allows for qubit manipulation \cite{girvin2009}.\\

 As for the ISJJ architecture, shown in \Fref{fig:circuits}(c), its Hamiltonian contains an additional term associated with the presence of the (linear) inductance {$L$}. It reads~\cite{girvin2009,vool2017}
\begin{equation}
	\label{eq:flux}
	H_{\rm IS} = - E_J \cos(\phi) + 4 E_C n^2 + \frac{1}{2} E_L (\phi-\phi_{\rm ext})^2,
\end{equation}
with $E_C = e^2/(2C_J)$ and $E_L = L^{-1} \Phi_0^2/(4\pi^2)$. The dimensionless parameter $\phi_{ext}$ takes into account the external flux piercing the loop defined by the two inductive elements of the circuit.\\

Despite the similar structure of the two Hamiltonians, the two circuits result in different, and dual, responses to (static) external knobs. In particular, the CSJJ is insensitive to external flux $\phi_{\rm ext}$ while its spectrum generally strongly depends on the offset charge $n_g$. The opposite holds for the ISJJ.
An instructive way to understand these complementary behaviors is to consider the unitary transformation $U=e^{i \phi_{\rm ext} n} e^{-i \phi n_g}$, which shifts the charge and the flux operators by
\begin{eqnarray}
	Un U^\dagger &= n + n_g \\
	U \phi U^\dagger &= \phi + \phi_{\rm ext}. 
\end{eqnarray}
Let us consider, at first, the presence of an external flux. While, in the CSJJ, its effect can be gauged away by $U$, this is not possible in the inductively shunted configuration, where $\phi_{\rm ext}$ physically describes a relative and gauge-independent shift between the branch fluxes of the two inductive elements.
As for the offset charge, its possible presence in the Hamiltonian $H_{\rm IS}$ can be readily gauged away by acting with $U$. The case of the CSJJ, however, requires special care because of its structure. Indeed, the presence of a floating and isolated superconducting island implies that its charge, expressed in terms of the number of Cooper pairs $n$, is an integer-valued operator. In complete analogy with mechanical systems where the discretization of momentum is associated with a position operator defined on a ring, the integer nature of $n$ is directly associated with the flux $\phi$ being a compact operator living on a circle. The wavefunctions of the systems are thus $2\pi$-periodic in the flux. As a consequence, even if the transformation $U$ can actually eliminate $n_g$ from the Hamiltonian $H_{\rm CS}$, it affects non-trivially the periodicity of the wavefunctions, i.e. $[U\Psi](\phi+2\pi)= e^{-i2\pi n_g} [U\Psi](\phi)$. As a result, the effect of $n_g$ in a CSJJ cannot be gauged away and, in general, it does affect the spectrum of the circuit\footnote{More precisely, since integer values of $n_g$ can always be gauged away without modifying the periodicity of the wavefunctions, it is the non-integer component of $n_g$ that affects the spectrum.}.\\ 

Regardless of their differences, both types of superconducting circuits can be seen as one-dimensional anharmonic quantum oscillators, featuring discrete low-energy spectra whose properties depend on the circuit parameters ($E_C, E_J, E_L$) as well as on the external knobs, $n_g$ or $\phi_{\rm ext}$. In order to define a qubit, one should assume excitations to higher states to be highly suppressed and exclusively focus on the two lowest energy levels, i.e. the ground state $|0\rangle$ and the first excited one $|1\rangle$. Therefore the Hamiltonian can be reduced into a two-dimensional subspace, whose corresponding two-level system description reads
\begin{equation}
	H_q = \frac{1}{2}\omega_q \sigma^z,
\end{equation}
where the parameter-dependent $\omega_q$ is the qubit frequency and $\sigma^z$ is the third Pauli matrix. A convenient representation of the state of a single qubit is provided by a unit sphere known as ``Bloch sphere"~\cite{nielsen_book, benenti_book}. In particular, every pure qubit state $|\psi\rangle = \alpha |0\rangle +  \beta |1\rangle$ correspond to a unit vector connecting the center of the sphere with a point on its surface. Adopting standard convention, the north (south) pole correspond to the state $|0\rangle$ ($|1\rangle$). The $z$-axis, representing the quantization axis of the qubit, is usually called the ``longitudinal axis". By contrast, the $x-$ and $y-$ axes are dubbed as ``transverse". In the rotating frame, i.e. applying the unitary operator $U_{rf} = e^{iH_q t}$ to the wavefunction, the Bloch vector is stationary and points at a specific point on the surface. If the qubit is a perfectly isolated system, the Bloch vector remains fixed and the information stored in the qubit does not get corrupted. However, possible interaction terms with the environment can dramatically affect the state of the qubit in several ways, as we discuss shortly.

\subsection{Qubit decoherence}
\label{sec:deco}
The various parameters and external knobs that control the qubit dynamics can feel stochastic fluctuations due to interactions with uncontrollable modes of the circuit environment. These fluctuations therefore affect the dynamics of the qubit, leading to loss of coherence and subsequent corruption of quantum information~\cite{weiss_book, schnirman2002}. The study of different noise sources in superconducting quantum circuits, such as the presence of non-equilibrium quasiparticles \cite{gustavsson2016,catelani2012} or impurities acting as additional two-level systems \cite{faoro2006, muller2019}, goes back to the early stage of qubit proposals and it is still a very active field of research~\cite{wendin2017, weiss_book, schnirman2002, clerk2010}. The microscopic understanding of the mechanisms responsible for decoherence is indeed extremely valuable in order to remove or, at least, to mitigate their effects at the hardware level ~\cite{oliver2013,paladino2014}. At the same time, a proper characterization of the noise ~\cite{papic2022,harper2020,wise2021}, is a prerequisite for the implementation of promising error-mitigation schemes acting at the software level \cite{temme2017,endo2021}. A  microscopic and comprehensive analysis of several noise sources is beyond the scope of the present Review. Here, we briefly sketch one of the most commonn framework used to quantify the coherence properties of a qubit and how these can be addressed in experiments. In particular, we describe the main qualitative effects that a generic noise source can have on qubit dynamics, in terms of longitudinal relaxation and pure dephasing. 
\subsubsection{Longitudinal relaxation}
\begin{figure}
	\centering
	\includegraphics[width=.7\textwidth]{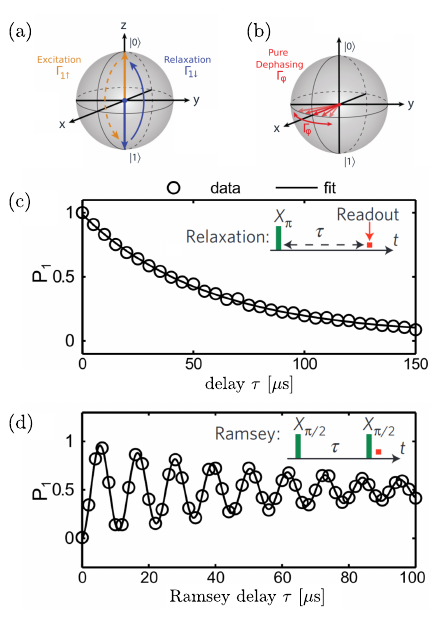}
		\caption{Representations of longitudinal relaxation (a) and pure dephasing (b) on the Bloch sphere \cite{krantz2019}. 
			Reproduced from \href{https://doi.org/10.1063/1.5089550}{Krantz et al., Applyed Physics Review \textbf{6} 021318 (2019)}, with the permission of AVS: Science \& Technology of Materials, Interfaces, and Processing.
			 Examples of measurements of longitudinal relaxation time $T_1 \sim 55 \,\mu$s (c) and coherence time $T_2^* \sim 56\, \mu$s (d) of a transmon qubit \cite{chang2013}. 
			 Reproduced from \href{https://doi.org/10.1063/1.4813269}{Chang et al., Applyed Physics Letters\textbf{ 103} 012602 (2013)}, with the permission of AVS: Science \& Technology of Materials, Interfaces, and Processing. 
			 The two insets sketch the measurement protocols (adapted from Ref.~\cite{bylander2011})}
	\label{fig:deco}
\end{figure}

The longitudinal relaxation describes the depolarization of the qubit along its quantization axis. It is determined by transverse noise, 
which allows for energy transfer between the qubit and the environment.
A prototypical interaction Hamiltonian responsible for longitudinal relaxation can be written as
\begin{equation}
	\label{eq:hrelax}
	H_{{\rm rel}} = \nu \sigma^x \otimes \lambda,
\end{equation}
where $\nu$ denotes the coupling strength and $\lambda$ represents a generic operator acting on the environment degrees of freedom and it is responsible for quantum fluctuations. This is an off-diagonal term in the qubit eigenbasis that induces transitions between the two eigenvalues of the qubit Hamiltonian $H_q$, $|0\rangle$ and $|1\rangle$. 
It is worth to remind that superconducting qubits typically operate at dilution refrigerators temperatures $T\sim 20\, {\rm mK} \sim 0.4\, {\rm GHz}$, whilst their characteristic frequency is in the range $5-10$GHz. 
Therefore, in this regime, induced transition from $|0\rangle$ to $|1\rangle$ are very unlikely to happens, and the interaction term in Eq.~\eref{eq:hrelax} is mostly responsible for decay $|1\rangle$ to the ground state $|0\rangle$, hence the name ``relaxation'' 
(see   \Fref{fig:deco}(a) for a pictorial view).\\

Being a resonant process, the longitudinal relaxation can only be induced by noise around the qubit frequency. Generally, for superconducting qubits, this kind of environmental interactions results in short correlation times and leads to an exponential decay\footnote{There are some specific exceptions to this statement, one example being the effects of quasiparticle poisoning, which typically leads to deviations from a simple exponential decay~\cite{gustavsson2016,pop2014}}, whose behavior can be captured by a so-called Bloch-Redfield picture \cite{weiss_book, breuer_book}. The time scale associated with this exponential decay is known as $T_1$. This can be estimated by evaluating the associated relaxation rate using for instance Fermi golden rule as
\begin{equation}
	1/T_1 = \Gamma_1 = \frac{1}{\hbar^2} \left|\langle 0 |
	\frac{\partial H_{{\rm rel}}}{\partial \lambda}
	| 1 \rangle
	\right|^2 S_\lambda(\omega_q),
\end{equation}
where $S_\lambda (\omega_q)$ is the power spectral density at the qubit frequency of the noise associated with fluctuations of the environment degree of freedom $\lambda$. A typical measurement scheme to address $T_1$ consists in preparing the qubit in the excited state $|1\rangle$ at $t=0$ and then letting it spontaneously decay. Projective measurements are then performed to determine the decay of qubit polarization as a function of time. An example of $T_1$ measurement, taken from Ref.~\cite{chang2013}, is provided in \Fref{fig:deco}(c).
\subsubsection{Pure dephasing}
A complementary interaction term describing the coupling with the environment is
\begin{equation}
	\label{hdephasing}
	H_{{\rm deph}} = \nu \sigma^z \otimes \lambda,
\end{equation}
where now the environmental degrees of freedom $\lambda$ are coupled to the longitudinal axis of the qubit along $\sigma^z$. While such a term does not induce transitions between energy levels of the qubit ($[H_{{\rm deph}}, H_q]=0$), it stochastically modulates the qubit frequency. The latter must be then considered as a function of one (or more) fluctuating variable(s) $\omega_q \to \omega_q(\lambda)$, where $\lambda$ can be a specific quantum operator of the environment [as in \Eref{eq:CPB}] or, more generally, a collection of external noisy parameters. Regardless of their specific origin, fluctuations of the qubit frequency determine a stochastic precession of the Bloch vector in the rotating frame, as depicted in \Fref{fig:deco}(b), and thus a gradual loss of phase coherence. The time scale associated with such a process is known as $T_\phi$ and it strongly depends on the qubit longitudinal sensitivity $\partial_\lambda \omega_q$. Within the Bloch-Redfield framework, meaningful for noise power spectral densities $S_\lambda(\omega)$ that are regular near $\omega \sim 0$ and up to $T_\phi^{-1}$, the pure dephasing time simply reads~\cite{ithier2005}
\begin{equation}
	\label{eq:Tphi}
	\frac{1}{T_\phi} = \pi \left|\frac{\partial \omega_q}{\partial \lambda}
	\right|^2 S_\lambda(\omega=0).
\end{equation}
While the presence of singularities in the noise spectrum leads to more cumbersome expressions for $T_\phi$, see e.g. Refs.~\cite{koch2007, ithier2005,martinis2003}, it is generally true that a reduction of longitudinal sensitivity $\partial_\lambda \omega_q$ (as well as of higher derivatives) leads to longer pure dephasing times.\\ 

In sharp contrast with longitudinal relaxation, pure dephasing is not a resonant process and it is instead determined by broadband noise. Moreover, since it does not involve energy transfer with the environment, pure dephasing is in principle reversible, i.e. it can be undone by applying proper unitary transformations. It is thus possible to devise dynamical decoupling protocols that dramatically increase $T_\phi$ by effectively filtering the noise (typically suppressing its low-frequency components)~\cite{paladino2014, bylander2011,uhrig2007,uys2009}. Notable examples are the Hahn echo scheme \cite{hahn1950} and the CPMG pulse sequence \cite{carr1954,meiboom1958}. Although direct access to $T_\phi$ is very hard to achieve, its value can be extracted by measuring the total decoherence time, as discussed below.    

\subsubsection{Decoherence}
The loss of coherence of a superposition state, for instance, $|\psi_i\rangle = (|0\rangle + e^{i\phi} |1\rangle)/\sqrt{2}$, is caused by the combined effect of pure dephasing (which leads to stochastic precession of $\phi$) and longitudinal relaxation (since, after a transition to $|0\rangle$, all the information about the original superposition will be lost). Its characteristic time, known as $T_2$, can be determined using Ramsey interferometry~\cite{ramsey1950}. Starting from the ground state $|0\rangle$, a $X_{\pi/2}$ pulse, i.e. a rotation of $\pi/2$ along the $x$-axes, brings the qubit to $|\psi_i\rangle = (|0\rangle + i |1\rangle)/\sqrt{2}$. A second $X_{\pi/2}$ pulse is then performed after a time $\tau$ \footnote{Typically, a controlled and deterministic precession around the $z$-axis is induced, in order to produce an oscillating Ramsey fringe, which is easier to analyze.}. While in the limit $\tau \to0$ the qubit reaches the excited state $|1\rangle$, for larger $\tau$, when decoherence kicks in, the ensemble-averaged polarization along the $z$-axis decays towards zero. An example of $T^*_2$ measurement, taken from Ref.~\cite{chang2013}, is provided in   \Fref{fig:deco}(d). Assuming an exponential shape for the decay functions, one obtains the following relation
\begin{equation}
	\frac{1}{T_2} = \frac{1}{2T_1} + \frac{1}{T_\phi},
\end{equation}
which in the end allow to determine $T_\phi$ from the experimental measurements of $T_1$ and $T_2$. It should be noted, however, that the broadband dephasing noise might not be well described within a Bloch-Redfield picture (for instance due to $1/f$-like components that exhibits singularities and long correlation times) and it can feature non-exponential behaviors, leading to deviations from the simple expression quoted above~\cite{bylander2011}\footnote{Note that even when the loss of coherence is not exponential, for sake of comparison among different decays, values of $T_2$ are typically determined by identifying the time associated with a decay factor of $1/e$~\cite{koch2007,bylander2011, ithier2005}.}.\\

As mentioned before, the pure dephasing contribution to decoherence can be reduced by implementing dynamical decoupling protocols in between the two $X_{\pi/2}$ pulses that are used for the measurements of $T_2$. Typically, an asterisk is used ($T_2^*$) to indicate the Ramsey coherence time, i.e. when no extra echo pulses have been used. By contrast, the (longer) coherence times obtained with the implementation of dynamically decoupling schemes is typically denoted by an additional ``e'' of ``echo'' ($T_{2e}$).\\ Finally, if $T_\phi$ is large enough, a $T_1$-limited regime is reached and the coherence time of the qubit is simply given by $T_2 = 2 T_1$. 

\subsection{State-of-the-art single-mode superconducting qubits}
We now briefly present two of the most widely investigated single-mode superconducting qubits, the transmon and the fluxonium qubit. They are based on the two single-mode circuits described in Section~\ref{sec:sup_quantum_circuits}, operating in specific regimes of the Hamiltonian parameters ($E_C$, $E_J$ and $E_L$). Precise tuning of these parameters significantly reduces the qubit sensitivity either to longitudinal or transverse noise. This point, which is a crucial ingredient for the great performance of state-of-the-art devices, is also extremely relevant for the development of the novel and more complex architectures, as we will discuss in the following. There, strong robustness against the noise stems indeed from the interplay between qubit design and peculiar parameter regimes. 

\subsubsection{Transmon qubit}
A transmon qubit relies on a CSJJ described by the circuit sketched in \Fref{fig:circuits}(b). It is modeled by the Hamiltonian 
\begin{equation}
	\label{eq:CPB_2}
	H_{\rm CS} = - E_J \cos(\phi) + 4E_C (n - n_g)^2,
\end{equation} 
already introduced in Eq.~\eref{eq:CPB}. Its peculiarity stems from its operational regime, which is characterized by $E_J \ll E_C$ \cite{krantz2019, girvin2009, koch2007,kjaergaard2020,roth2022}. 
This regime of energy scales can be achieved by using a large shunt capacitance in order to lower $E_C$, typically reaching ratios of the order of $E_J/E_C \sim 50-60 $. First realizations of transmon qubits were based on interdigitated capacitors~\cite{koch2007,weides2011} while, more recently, different geometries are typically favored, like cross-shaped capacitors in the so-called Xmon design \cite{barends2013}, in order to achieve better interconnectivity and scalability.\\

In addition to the insensitivity to low-frequency flux noise, featured by all CSJJ, the regime $E_J \gg E_C$ guarantees exponential protection against pure dephasing induced by fluctuations of the offset charge $n_g$. Indeed, for large ratios $E_J/E_C$, the charge dispersion flattens for all values of $n_g$, leading to an exponential suppression of the derivative ~\cite{koch2007}
\begin{equation}
	\label{eq:partialW_partialNG}
	\frac{\partial \omega_{01}}{\partial n_g} \propto e^{-\sqrt{8 E_J/E_C}}
\end{equation}
and therefore, according to Eq.~\eref{eq:Tphi}, of the charge noise-induced dephasing. A detailed demonstration of Eq.~\eref{eq:partialW_partialNG} can be found, for example, in Refs.~\cite{girvin2009, koch2007}. However, it is important to understand the physical origin of the charge dispersion flattening. When $E_J$ dominates over $E_C$, the Hamiltonian $H_{\rm CS}$ is characterized by a very deep potential well, centered at $\phi = 0$. The wavefunctions of the lowest energy states are therefore well localized around $\phi \sim 0$, meaning that they are extremely widespread when expressed in terms of the conjugated variable $n$, extending over several integer values of the charge. Variations of the offset charge are therefore averaged out and the spectrum becomes almost independent of $n_g$.\\ 

The (exponential) protection against charge noise-induced pure dephasing dramatically improves the coherence of the transmon qubit. This, combined with high technological maturity and good scalability, explains why most of the largest and most powerful quantum processors available to date are indeed based on such a qubit design~\cite{arute2019,IBM2022}. Transmon qubits, however, also possess drawbacks, some of them directly related to the very presence of a deep potential well in the Hamiltonian, which might limit further developments. Indeed, at the bottom of the cosine, the potential is almost quadratic, which leads to a small relative anharmonicity (typically around $5\%$ of the qubit frequency). This makes it more complicated to prevent leakages from the computational space to higher energy states. Luckily, the loss of anharmonicity is only linear in the ratio $\sqrt{E_J/E_C}$ and the exponential gain in terms of coherence still makes up for it~\cite{koch2007, roth2022}. Another important issue is related to the relaxation time $T_1$, which does not benefit from any kind of protection. As a matter of fact, the strong wavefunction localization around the minimum of the potential well necessarily implies a large wavefunction overlap for the states $|0\rangle$ and $|1\rangle$, actually favoring energy relaxation processes. Finally, we underline that charge noise is not the only source of dephasing, which can originate also from other mechanisms lacking exponential protection, such as fluctuations of $E_J$ induced by spurious localized two-level systems embedded in the JJs (the so-called critical current noise)~\cite{martinis2003,koch2007,papic2022,muller2019,nugroho2013} and flux noise in the (commonly used) tunable split-transmon configuration \cite{krantz2019,koch2007, kjaergaard2020}. 

\subsubsection{Fluxonium qubit}
\label{sec:fluxonium}
The optimization of qubit designs and operating parameter regimes, aiming at reducing noise sensitivity, cannot ignore a fundamental relation between charge and flux noise, stemming from Heisenberg's uncertainty principle. Large flux fluctuations $\delta \phi$ are indeed necessarily associated with (conjugated) low charge fluctuations $\delta n$ and vice versa. The physical quantity that determines whether a system features large charge or flux fluctuations (``large'' in terms of their fundamental units) is the circuit impedance $Z$, usually expressed in units of the superconducting quantum of resistance $R_Q = h/(2e)^2 = 6.5\, {\rm k}\Omega$~\cite{manucharyan2012}. Considering the paradigmatic example of a simple harmonic LC oscillator, one can indeed show that\footnote{Fluctuations of charge (flux) contributes to the total energy of the LC circuit as $4 e^2 (\delta n)^2/(2C)$ ($\Phi_0^2 (\delta \phi)^2/(8 \pi^2L)$). By exploiting the virial theorem, which ensures that both these energies are equal, it is straightforward to obtain Eq.~\eref{eq:noise_imp} \cite{manucharyan2012}.} 
\begin{equation}
	\label{eq:noise_imp}
	\frac{\delta \phi/(2\pi)}{\delta n} = \frac{Z}{R_Q},
\end{equation}
where $Z = \sqrt{L/C}$ is the impedance of the LC circuit. Notably, the electromagnetic environments of the qubits are typically characterized by low impedance values. Indeed, in presence of geometric inductances and capacitances, whose properties ultimately originate from magnetic and electric fields stored within the system, impedances are \red{typically} limited by the vacuum impedance $Z_{\rm vac} =\sqrt{\mu_0/\epsilon_0} \sim 377\Omega \ll R_Q$ ($\epsilon_0$ and $\mu_0$ being the vacuum permeability and permittivity, respectively). Notice that substrate materials with large dielectric constants (above $10$ for SiO$_2$) further reduce typical impedance values, which therefore lay well in the \textit{low} regime \cite{manucharyan2012}. This leads to charge noise typically dominating over flux noise \cite{manucharyan2012}.\\  

This observation justifies the interest in ISJJ circuits, shown in \Fref{fig:circuits}(b), whose spectrum is intrinsically insensitive to charge noise. Moreover, in complete analogy with the transmon qubit, one could even devise a parameter regime that guarantees exponentially suppressed sensitivity also with respect to flux noise~\cite{koch2009}. In such a regime, corresponding to $E_C\sim E_J$ and $E_J/E_L \gtrsim 100$~\cite{pechenezhskiy2020}, the wavefunctions of the computational states are indeed spread over several local minima of the cosine potential, making low-frequency fluctuations of the external flux $\phi_{\rm ext}$ harmless. This scenario, which has been recently experimentally realized~\cite{pechenezhskiy2020}, is however less than ideal, as it requires extremely large kinetic inductance, obtained with superinductors (that are experimentally challenging and with scalability issues), and shares transmon limitations (small anharmonicity and absence of protection against energy relaxation processes). 

In this respect, the fluxonium qubit~\cite{manucharyan2009}, represents a good compromise as it operates in the more moderate regime $E_J/E_L\sim 3-20$ and $E_J/E_C \gtrsim 1$~\cite{manucharyan2009,somoroff2021, lin2018, nguyen2019}. In this case, the spectrum does depend on the external flux but, at the sweet spot $\phi_{\rm ext} = \pi$, the first derivative $\partial_{\phi_{\rm ext}} \omega_{q}$ vanishes. This still guarantees first-order protection against flux noise, which is typically already enough given the small amplitude of flux fluctuations. At the same time, the fluxonium retains a large anharmonicity and a rich structure of the excited states (which allows for convenient exploitation of non-computational transitions to design on-demand multi-qubit interactions \cite{nesterov2018, nesterov2022,chen2022}). The wavefunction overlap between states $|0\rangle$ and $|1\rangle$ is also reduced with respect to the transmon scenario~\cite{somoroff2021}, potentially leading to longer relaxation times $T_1$. One of the main limitations of the fluxonium qubit is that its implementation necessarily requires superinductors, \blue{in order to have large enough impedances.} As we will review in Section~\ref{sec:superinductor}, the fabrication of superinductors is not straightforward, representing an open and active field of research, and their complex structure can even introduce some degree of sensitivity to charge noise (via phase slip events). Because of all these aspects, fluxonium qubits have the potential to soon become an alternative and competing platform with respect to transmons \cite{bao2021, nguyen2022}. 
\subsubsection{Performance of state-of-the-art qubits}
The first decade of the new millennium has witnessed an improvement in the performance of superconducting qubits by order of magnitudes, starting from coherence times in the realm of nanoseconds featured by the first realizations of charge-box qubits~\cite{nakamura1999, nakamura2002}, all the way to tens of microseconds, for example in 2D transmon qubits~\cite{Houck2008,chow2012,chang2013}. After a period characterized by a somehow slower \red{pace} in the improvement of coherence times~\cite{kjaergaard2020, nersisyan2019}, recent breakthroughs involving new materials and qubit designs eventually allowed for notable leaps over the last few years. For instance, tantalum-based transmons have recently reached relaxation times of the order of $T_1\gtrsim 0.3\, ms$ \cite{place2021} and $T_1\sim 0.5\, ms$ \cite{wang2021}. As for fluxonium qubits, the recent realization of high-coherence devices allowed to reach $T_{2e} > 100\mu s$ in 2019 \cite{nguyen2019}, $T_{2e} \sim 300\mu s$ in 2021 \cite{zhang2021} and, finally, to enter the realm of {milliseconds} Ramsey coherence times $T_2^* = 1.48\, ms$ \cite{somoroff2021}.\\

We conclude this Section by mentioning that the coherence time of a qubit, despite being a very important metric, is far from being enough to fully characterize the performance of a given quantum processor. While a detailed discussion of error benchmarking is well beyond the scope of this Review (the interested readers may consult a number of publications in the literature such as Refs.~\cite{wootton2020, cross2016,gaebler2012, mcKay2017}), it is useful to briefly comment on the broader problem of assessing the overall quality of a quantum device. The most obvious reason why the knowledge of $T_1$ and $T_2$ is not enough is that they must be compared with the typical execution times of single- and two-qubit gates. The latter depend on several factors such as the qubit anharmonicity, which sets an upper limit to the gate execution speed. In this respect, the exploration of parameter regimes that maximizes the qubit anharmonicity (see, e.g., Ref. \cite{hyyppa2022}) might be as important as the increase of coherence times. A typical example of local metrics that naturally take into account the actual implementation of quantum gates is represented by the single- and two-qubit gate fidelities, which measure deviations between the ideal and observed output of quantum gates \cite{magesan2012}. Superconducting qubits have recently reached very high single-qubit gate fidelities, with transmons reaching $0.9998$ \cite{mcKay2017} and fluxonium qubits recently exceeding the $0.9999$ threshold \cite{somoroff2021}. As for two-qubit gates, the best transmon and fluxonium-based setups reach fidelities above $0.99$ \cite{bao2021,sung2021, ficheux2021,kandala2021}. Importantly, even local metrics like gate fidelities are not enough to properly characterize complex multi-qubit devices, which can feature, for example, significant spatial and temporal variations in the properties of individual qubits. Even though a number of multi-qubit metrics have been proposed and measured \cite{flammia2011,dasilva2011,mcKay2019,proctor2019,cross2016,wei2020}, ranging from entanglement-based metrics to quantum volume, how to best benchmark a quantum device is still an open and important question \cite{preskill2018}.  

\section{Multi-mode superconducting qubits}
\label{sec:multimode}
Let us summarize some important concepts introduced in the previous Section. We have seen that qubit decoherence results from the combination of energy relaxation and pure dephasing mechanisms. To reduce the former, one has to suppress the transition amplitude $|\langle 0|H_{\rm rel}|1\rangle|^2$ between the computational states [see Eq.~\eref{eq:hrelax}]. Since the Hamiltonian $H_{\rm rel}$ that couples the qubit to the environment is typically local in charge or flux space, minimizing the corresponding spatial overlap between the qubit wavefunctions generally guarantees a strong suppression of relaxation-induced decoherence. By contrast, pure dephasing depends on the sensitivity $|\partial_\lambda \omega_{q}|$ of the qubit frequency $\omega_{q}$ with respect to some fluctuating parameters $\lambda$. If the latter is an external flux (an offset charge), a convenient strategy to suppress dephasing is to strongly delocalize the qubit wavefunctions in flux (charge) space and, correspondingly, localize them in the dual charge (flux) space.
In simple single-mode circuits, such as transmon and fluxonium qubits, it is not possible to simultaneously suppress these two decoherence mechanisms. There, a strong localization of wavefunctions, engineered to reduce pure dephasing, is indeed necessarily associated with a large wavefunction overlap, which is highly detrimental for the suppression of energy relaxation.\\

\red{This} triggered the development of more complicated qubit architectures, featuring richer Hamiltonians that can support localized eigenstates with negligible overlap. Such a desirable scenario, \red{an important ingredient toward} the realization of fully protected qubits, can be engineered in systems with more than one degree of freedom and/or characterized by the presence of specific symmetries. This is at the basis of so-called multi-mode qubits. Over the last two decades, several setups of this kind have been theoretically proposed and experimentally investigated~\cite{doucot2012, gyenis2021, dannon2022, ioffe2002, doucot2002, kitaev2006, brooks2013, smith2020, gladchenko2009, bell2014, pechenezhskiy2020, kalashnikov2020, gyenis2021_exp}. In order to explain the working principle of multi-mode qubits, and the physical reasons that lead to full noise protection, in this Section we describe in detail the paradigmatic proposal of the so-called $0-\pi$ circuit, introduced for the first time by Brooks and collaborators in Ref.~\cite{brooks2013} and directly inspired by the ideas laid out in Ref.~\cite{kitaev2006}.
Its theoretical description is indeed ideal for pedagogical purposes, as it allows us to introduce at once several key concepts which applies also to generic multi-mode architectures.\\

\blue{The noise protection offered by these advanced designs comes at the price of a more complicated experimental realization and non-trivial qubit manipulation. The challenges associated with the fabrication of the circuits are discussed in the next Section, where we review several key experiments and highlight the differences between optimal parameter regimes and the ones that have been actually achieved in the lab. As for qubit manipulations, the main issue is that the very protection from environmental noise makes it quite complicated to actively control the device by applying quantum gates. In \Sref{sec:gates}, we discuss this important point by reviewing some recent theoretical proposals for the execution of specific gates, a formidable task that still represents an important and active research areas. Despite these challenges, the ongoing development of multi-mode qubits has the potential to boost the performance of superconducting-based quantum computation, for example by exploiting hybrid architectures featuring both single-mode and multi-mode qubits \cite{maiani2021, ciani2022}.
}

\subsection{The $0-\pi$ circuit}
\begin{figure}
	\centering
	\includegraphics[width=0.7\linewidth]{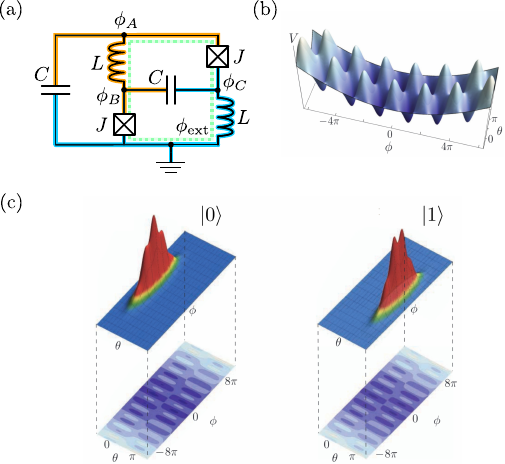}
	\caption{(a) Electrical circuit of the symmetric $0-\pi$ circuit. The circuital parameters are indicated with $C$, $L$ and $J\equiv (E_J, C_J)$. The three active nodes have node fluxes $\phi_A$, $\phi_B$, $\phi_C$. The circuit features two superconducting islands, highlighted in blue and orange, as well as one inductive loop (dashed green line) pierced by the external flux $\phi_{{\rm ext}}$. (b) Potential landscape $V(\theta,\phi)$ of Hamiltonian $H_{\theta\phi}$, given in \Eref{eq:V}, in the deep $0-\pi$ regime. (c) Logical wavefunctions of the $0-\pi$ qubit in the deep regime. The ground state $|0\rangle$ is sharply localized on the $\theta \sim 0$ ridge. By contrast, the first excited state (which is almost degenerate with the ground state) is sharply localized on the $\theta \sim \pi$ ridge. The spatial overlaps of the two wavefunctions is exponentially small.		
		 Panels (b) and (c) adapted from Ref.~\cite{dempster2014}.}
	\label{fig:0pi_circuit}
\end{figure}

The superconducting circuit that allows realizing the $0-\pi$ qubit has been already depicted in \Fref{fig:circuits}(d), where its richer multi-mode structure can be directly compared to the simpler ones featured by single-mode CSJJ and ISJJ circuits [panels (b) and (c)]. The circuit, reported also in \Fref{fig:0pi_circuit}(a), consists of a pair of JJs, a pair of linear inductors, and a pair of cross capacitances. It features four nodes, one of which can be chosen as a reference by setting its node flux to zero. The system has thus a total of three degrees of freedom, which can be identified with the three-node fluxes $\phi_A$, $\phi_B$, and $\phi_C$ shown in \Fref{fig:0pi_circuit}(a).
A quick analysis of the circuit topology shows that it shares some important features with the CSJJ and ISJJ architectures. In particular, there are two isolated superconducting islands coupled by Josephson elements [see the blue and orange regions], like in the CSJJ architecture, and an inductive loop pierced by an external flux $\phi_{\rm ext}$ [see the dashed green rectangle], in analogy with the ISJJ circuit. These similarities will allow us to leverage the concepts introduced in the previous Section to readily understand several properties of the $0-\pi$ qubit, which can be indeed interpreted as an ingenious combination of a transmon and a fluxonium qubit.\\ 

At first, let us derive the Hamiltonian of the circuit in the symmetric case, i.e. when circuital elements are pairwise identical. In this ideal scenario, the circuit is completely described by four parameters: The Josephson energy $E_J$ and self-capacitance $C_J$ of the JJs, the inductance $L$, and the capacity $C$. The linear elements of the circuit, i.e. everything but the two Josephson elements, are described by the quadratic Lagrangian
\begin{equation}
	\mathcal{L}^Q = \dot {\Sigma}^T \bar C \dot \Sigma - \Sigma^T \bar L {\Sigma},
\end{equation}
where $\Sigma = (\phi_A, \phi_B,\phi_C)^T$ is the vector of node fluxes. The capacitance and inductance matrices read
\begin{eqnarray}
	\bar C &= \frac{1}{2} \left(\begin{array}{*{20}{c}}
		C+C_J& 0& -C_J\\
		0& C+C_J & -C \\
		-C_J & -C & C+C_J 
	\end{array}\right)\\
\bar L &= \frac{1}{2}  \left(\begin{array}{*{20}{c}}
		L& -L& 0\\
		-L& L & 0 \\
		0 & 0 & L \\
	\end{array}\right).
\end{eqnarray}
The normal mode of the linear circuit, which can be obtained by diagonalizing $\bar C^{-1}\bar L$, read~\cite{dempster2014}
\begin{eqnarray}
	\label{eq:thetaNM}
	2\theta &= +\phi_A + \phi_B - \phi_C \\
	\label{eq:xi}
	2\xi &= + \phi_A - \phi_B + \phi_C \\
	\label{eq:phiMN}
	2\phi &= - \phi_A + \phi_B + \phi_C
\end{eqnarray}
and allow  to conveniently express the Lagrangian of the whole circuit as 
\begin{eqnarray}
	\label{eq:Lsym}
	\mathcal{L^{\rm sym}} & =2E_J \cos(\theta)\cos(\phi) \\&- \left(\frac{\Phi_0}{2\pi}\right)^2 \left[\frac{1}{L} (\phi-\phi_{\rm ext}/2 )^2 + \frac{1}{L} (\xi-\phi_{\rm ext}/2 )^2\right] \\
	& + \left(\frac{\Phi_0}{2\pi}\right)^2 \left[ C \dot\theta^2+ C \dot\xi^2 + C_J \dot\theta^2 + C_J \dot\phi^2 \right].
\end{eqnarray}
The Legendre transform in \Eref{eq:Legendre}, followed by canonical quantization, lead then to the quantum Hamiltonian $H^{\rm sym} = H_{\theta\phi} + H_\xi$ with~\cite{dempster2014}
\begin{equation}
	\label{eq:Hsym}
	H_{\theta\phi} = 2E_{C}^{\theta} n_\theta^2 + 2E_{C_\phi} n_\phi^2 + E_L \phi^2 - 2E_J \cos(\theta)\cos(\phi+\phi_{\rm ext}/2),
\end{equation}
and the harmonic term 
\begin{equation}
	\label{eq:Hxi}
	H_\xi = 2E_{C}^{\xi} n_\xi^2 + E_L (\xi-\phi_{\rm ext}/2)^2.
\end{equation}
Here we have introduced $E_{L}=L^{-1} \Phi_0^2/(4\pi^2)$ and the charging energies $E_{C}^{\xi} = e^2/(2C)$, $E_{C}^{\phi}=e^2/(2C_J)$, $E_{C}^{\theta}=e^2/(2C+2C_J)$. The charge operators $n_{\xi}, n_\phi,$ and $n_\theta$ are conjugated to the corresponding flux operators, i.e. $[\xi,n_\xi] = [\phi,n_\phi] = [\theta,n_\theta] = i$. For the sake of convenience, we have also performed a gauge transformation $e^{in_\phi\phi_{{\rm ext}}/2}$ to move the external flux into the cosine term.\\

It is worth to notice that the $\xi$ mode does not bias the JJs\footnote{In order to see this point more clearly, it can be useful to change the gauge and assign a finite value to the node flux of the reference node, which we call $\phi_0$. The definition of the $\xi$ mode in \Eref{eq:xi} must be then expressed in terms of the {differences} between the node fluxes and reads
	\begin{equation}
		2\xi = (\phi_A-\phi_0) - (\phi_B-\phi_0) + (\phi_C-\phi_0) = (\phi_A +\phi_C)-(\phi_B+\phi_0).
	\end{equation}
	This formulation makes it more evident that $\xi$ is not influenced by the phase drops on the JJ.}. It is therefore purely harmonic and completely decoupled from the other two modes, $\theta$, and $\phi$. This fact, which holds true {only} for ideal symmetric circuits, greatly simplifies the description of the $0-\pi$ circuit. Indeed, it allows to focus only on the effective Hamiltonian $H_{\theta\phi}$. The latter is intrinsically two-dimensional since the Josephson term couples in a non-linear way the two modes $\phi$ and $\theta$. Apart from this important point, however, $H_{\theta\phi}$ can be seen as a combination of $H_{\rm CS}$ and $H_{\rm IS}$, i.e. the two single-mode Hamiltonians that describe the CSJJ and ISJJ architectures, respectively [see Equations \eref{eq:CPB} and \eref{eq:flux}]. This observation, in complete agreement with the circuit topology [see \Fref{fig:0pi_circuit}(a)], will allow us to readily derive some important coherence properties of the $0-\pi$ qubit, the latter being defined by the two lowest energy levels of $H_{\theta\phi}$.  

\subsection{Noise protection in the deep $0-\pi$ regime}
The $0-\pi$ qubit features the highest degree of noise protection when operated in the so-called \textit{deep regime} \cite{brooks2013,dempster2014,groszkowski2018,diPaolo2019}, which corresponds to the constraints 
\begin{equation}
	\label{eq:deep}
	\left\{
	\begin{array}{*{20}{l}}
		E_L/E_J  \gg 1\\
		E_{C}^{\theta} /E_J  \gg 1\\
		E_{C}^{\phi} /E_J\sim 1
	\end{array}\right..
\end{equation}
This regime guarantees indeed simultaneous and exponential protection against pure dephasing, caused by charge and flux noise, and energy relaxation.\\

Protection against dephasing originates from the very same mechanisms exploited by transmon and fluxonium qubits and analyzed in the previous Section. In particular, akin to the single-mode of a transmon qubit, the $\theta$ mode is intrinsically insensitive to flux noise (since shifts of $\theta$ can be gauged away) and a large ratio $E_J/E_{C}^{\theta}$ exponentially suppresses the sensitivity to charge noise by reducing the charge dispersion~\cite{koch2007}.
 As for the $\phi$ mode, in analogy with the single-mode  fluxonium qubit, it is intrinsically insensitive to charge noise while its external flux sensitivity can be significantly reduced in the fluxonium regime, i.e. $E_J/E_{C}^{\phi} \gtrsim 1$ and $E_J/E_L \sim 3-20$, or even exponentially suppressed for even larger values of $E_J/E_L$~\cite{koch2009,pechenezhskiy2020}.
Conversely, the protection against relaxation is a unique consequence of the peculiar structure of the 2D Hamiltonian $H_{\theta\phi}$ and has no counterpart in the simple single-mode architectures. To develop some intuition, following Ref.~\cite{dempster2014}, it is instructive to study the 2D potential landscape of the Hamiltonian $H_{\theta\phi}$, which reads 
\begin{equation}
	\label{eq:V}
	V(\theta,\phi,\phi_{{\rm ext}}) = E_L \phi^2 - 2E_J \cos(\theta)\cos(\phi+\phi_{\rm ext}/2)
\end{equation}
and it is shown in \Fref{fig:0pi}(b). It consists of two ridges, centered around $\theta=0$ and $\theta=\pi$, each one featuring several minima in the $\phi$ direction and a very shallow overall parabolic envelope, associated with the small $E_L$ term. Each set of minima is staggered with respect to the other one. In the deep regime, the two wavefunctions associated with the computational eigenstates $|0\rangle$ and $|1\rangle$ are highly delocalized in the $\phi$ direction (fluxonium-like regime) and strongly localized in the $\theta$ space (transmon-like regime). Differently from the case of the transmon, however, the existence of two distinct ridges allows for the localization of each logical wavefunction around a different value of $\theta$. In particular, as shown in \Fref{fig:0pi_circuit}(c), the state $|0\rangle$ is localized around $\theta \sim 0$ while $|1\rangle$ is localized around $\theta \sim \pi$. This fact, from which the $0-\pi$ qubit takes its name, strongly suppresses the spatial overlap of the wavefunctions and thus the relaxation induced by local perturbations.\\ 

An additional and intriguing feature of the deep $0-\pi$ regime is that the two computational states have almost the same energy, up to exponentially small deviations. This might come as a surprise since the staggered nature of the potential makes the wavefunction localized on the $\theta \sim 0$ ridge to experience minima at different positions (in the $\phi$ direction) than the other wavefunction localized around $\theta \sim \pi$ [see \Fref{fig:0pi_circuit}(c)]. As a consequence, with the exception of the sweet spot $\phi_{{\rm ext}} = \pi$, when the potential features the inversion symmetry $V(\theta,\phi,\phi_{{\rm ext}} =\pi) = V(\theta+\pi,-\phi,\phi_{{\rm ext}} =\pi)$, the two computational states are non-degenerate. In the deep regime, however, the delocalization of the wavefunctions along the $\phi$ direction makes their energy almost insensitive to the exact position of the minima, leading to an (almost) degenerate computational manifold. This offers additional protection against other dephasing mechanisms, such as fluctuation of $E_J$ due to critical current noise~\cite{groszkowski2018}.

\subsection{Current mirror and effective single-mode description}
\label{sec:1deff}
The properties of the $0-\pi$ circuit can also be understood as the consequence of an intriguing physical phenomenon that characterizes the deep parameter regime: The dominance of co-tunneling of Cooper pairs (i.e. two correlated tunneling events, one on each JJ) over single tunneling events (involving only one Cooper pair and one JJ). The connection between such a general physical picture and the protection of the qubit extends well beyond the specific $0-\pi$ circuit considered so far. As a matter of fact, it holds for several other protected setup and often provides a useful and quick way to develop an intuition about their properties. It is therefore important to highlight its key ingredients.\\ 

\begin{figure}
	\centering
	\includegraphics[width=0.6\linewidth]{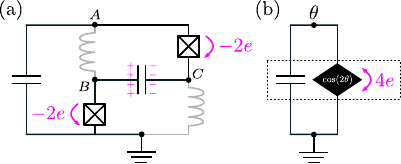}
	\caption{(a) Sketch of the current mirroring effect that justifies the predominance of Cooper pair co-tunneling. On a short time scale, the large inductors (in gray) strongly oppose on any change in the current flow (due to their large impedance). To rapidly equilibrate the charge imbalance resulting from the tunneling of a single Cooper pair on a JJ, a second correlated tunneling event must occur on the other JJ. (b) Effective single-mode circuit consisting of two superconducting islands that can only exchange pair of Cooper pairs.  }
	\label{fig:1Deff}
\end{figure}

Following Ref.~\cite{diPaolo2019}, we motivate the co-tunneling picture by considering the effects of a tunneling of a single Cooper pair from node A to node C, across the corresponding JJ [see the purple labels in \Fref{fig:1Deff}(a)]. On a short time scale, the large inductors don't allow any change in current flow (as they have high impedance), leading to a negative charge build-up on the left side of the large capacitor connected to node C. This must be compensated by a positive charge on the other side of the capacitor, i.e. on node B. The only way to quickly provide such a charge is for a second Cooper pair to tunnel from B to the reference node, across the second JJ. As a result, the peculiar design of the $0-\pi$ circuit strongly suppresses the isolated tunneling of a Cooper pair, as the latter tends to be immediately mirrored by a second tunneling event. This concept of current mirroring, crucial for the development of several protected qubits, has been introduced for the first time by A. Kitaev in Ref.~\cite{kitaev2006}.
Based on the previous discussion, one might be tempted to completely eliminate the two large inductors, which were indeed not considered in the setups analyzed in Ref.~\cite{kitaev2006}. Their presence, however, is crucial for the $0-\pi$ circuit, as it allows to define only two superconducting islands (as opposed to four) and guarantees slow charge equilibration within each one of them. As a result, the slow dynamics of the whole circuit can be simply described in terms of the tunneling of pairs of Cooper pairs between two superconducting islands. In complete analogy with the CSJJ architecture, such a scenario can be thus described by a single-mode effective Hamiltonian 
\begin{equation}
	\label{eq:H1Deff}
	H^{\rm 1D}_{\rm eff} = 2E_{C}^{\theta} n_\theta^2 - E_2\cos(2\theta),
\end{equation}
consisting of a generic charging energy term and a non-linear inductive one, featuring $\pi$ periodicity in the flux variable $\theta$ instead of $2\pi$, since it describes the tunneling of two Cooper pairs at a time \cite{brooks2013}. This is schematically depicted in \Fref{fig:1Deff}(b) and a qualitative sketch of its spectrum can be found in \Fref{fig:Gates}(e). The parity $P$ of the number of Cooper pairs, given by $P=(-1)^{n_\theta} = e^{-i n_\theta \pi}$, is thus a conserved quantity. Correspondingly, the operator $P$, whose action on the flux operator is a $\pi$ shift, i.e. $P \theta P^\dagger  = \theta - \pi$, is a symmetry of the $H^{\rm 1D}_{\rm eff}$. Such a simple Hamiltonian can be used to effectively describe several protected setups, which we will refer to as $\pi$-periodic qubits.\\ 

Before inspecting more properties of $H^{\rm 1D}_{\rm eff}$, it is instructive to complement the qualitative reasoning outlined before with a more rigorous derivation of $H^{\rm 1D}_{\rm eff}$. This can be done directly from the 2D Hamiltonian $H_{\theta\phi}$ of the $0-\pi$ circuit, by means of a Born-Oppenheimer approximation. To this end, we focus firstly on the fast and less massive degree of freedom $\phi$, whose mass term $(4E_{C}^{\phi})^{-1}$ is indeed much lighter than $(4E_{C}^{\theta})^{-1}$, the mass term of the slow and heavy mode $\theta$. The latter can be therefore treated as a fixed parameter $\bar \theta$ while solving the 1D Hamiltonian 
\begin{equation}
	H_\phi = 2E_{C}^{\phi} n_\phi^2 + E_L\phi^2 -2E_J \cos(\bar \theta) \cos(\phi+\phi_{\rm ext}/2).
\end{equation}
The ground state energy $E(\bar \theta,\phi_{\rm ext})$, which can be obtained by numerically diagonalizing $H_\phi$, represents then an effective potential energy for the slow $\theta$ mode. The dynamics of the latter can be then described very accurately by the effective 1D Hamiltonian \cite{diPaolo2019}
\begin{equation}
	\label{eq:2theta}
	H_\theta = 2E_{C}^{\theta} n_\theta^2 - E_2(\phi_{\rm ext})\cos(2\theta)-E_1 (\phi_{\rm ext})\cos(\theta).
\end{equation}
As expected from the previous considerations, this Hamiltonian contains a charging energy term and a $\pi$-periodic Josephson term $\cos(2\theta)$, describing the tunneling of pairs of Cooper pairs ($e^{-i2\theta} n_\theta e^{i2\theta} = n_\theta+2$). The third term, $\cos(\theta)$, is associated with the standard tunneling of a single Cooper pair, and its amplitude is exponentially suppressed in the deep $0-\pi$ regime. In particular, in Ref.~\cite{diPaolo2019} it is shown that 
\begin{eqnarray}
	E_2(\phi_{\rm ext}) &= E_\alpha - E_\beta \cos(\phi_{ext}),\\
	\label{eq:E1}
	E_1(\phi_{\rm ext}) &= E_\gamma \cos(\phi_{ext}/2),
\end{eqnarray}
where $E_\alpha/E_J$ is almost constant, while $E_\beta/E_J$ and $E_\gamma/E$ are exponentially small in the ratio $E_{C}^{\phi}/E_L$. We can therefore consider $H_\theta \sim H^{\rm 1D}_{\rm eff}$ and $P$ to be an approximate symmetry also of $H_\theta$.\\ 

The noise protection of the $0-\pi$ qubit, previously explained by analyzing the 2D Hamiltonian $H_{\theta\phi}$, can be also understood from the properties of the effective 1D Hamiltonian $H^{\rm 1D}_{\rm eff}$. Indeed, similarly to the CSJJ Hamiltonian, $H^{\rm 1D}_{\rm eff}$ is intrinsically insensitive to flux noise and, in the deep regime $E_{C}^{\theta}\ll E_2\sim E_J$ the sensitivity to charge noise is exponentially suppressed because of the strong localization of the logical wavefunctions in the wells of the cosine. The $\pi$ periodicity guarantees the existence of two distinct minima, so that the logical states $|0\rangle$ and $|1\rangle \simeq P|0\rangle$ are degenerate and with negligible overlap since one is localized around $\theta \sim 0$ and the other one around $\theta \sim \pi$. This  degeneracy can be lifted by a standard $2\pi$-periodic potential, like the $E_1$ term in $H_\theta$, that accounts for conventional single Cooper pair tunneling. \Eref{eq:E1}, however, shows that such a term vanishes for $\phi_{{\rm ext}}=\pi$, i.e. the same flux sweet-spot we have highlighted before when discussing the shape of the 2D potential landscape. In the deep regime, the $E_1$ is exponentially suppressed, regardless of the specific value of the external flux, and the system is thus always (almost) degenerate.

\subsection{Effects of spurious modes and overall coherence properties}
\label{sec:disorder}
So far we have considered an ideal symmetric circuit, where all the circuital elements are pairwise identical. Because of unavoidable fabrication imperfections, however, actual implementations of the $0-\pi$ circuit typically feature some amount of disorder in their parameters. As we will see, asymmetries between the circuital elements can induce a finite coupling between the harmonic mode $\xi$, introduced in Equations \eref{eq:xi} and \eref{eq:Hxi}, and the two degrees of freedom of the ideal $0-\pi$ circuit, the $\theta$ and $\phi$ modes. Even though the detrimental effects of this spurious coupling can be suppressed in the deep regime, they still represent one of the most important sources of decoherence of the $0-\pi$ circuit \cite{groszkowski2018}.\\
 
A convenient way to take into account the presence of disorder is to introduce the parameter averages $\bar Y = (Y_1+Y_2)/2$ and the relative deviations $\delta Y = (Y_1-Y_2)/Y$, with $Y\in\{E_J, E_L, C, C_J\}$. The Lagrangian of the whole circuit can be then rewritten as
\begin{eqnarray}
	\mathcal{L}  &=\mathcal{L}^{\rm sym} - E_J\delta E_J \sin(\theta)\sin(\phi)\\
	&+\left(\frac{\Phi_0}{2\pi}
	\right)^2 \left[\frac{1}{2}C \delta C \; \dot{\theta}  \dot \xi + \frac{1}{2}C_J \delta C_J \; \dot{\theta}  \dot \phi - \frac{1}{2}L \delta L \; {\xi}   \phi  
	\right].
\end{eqnarray}
By assuming the relative capacitance deviations $\delta C$ and $\delta C_J$ to be small, we derive the new expressions for the conjugated charges $n_\theta$, $n_\phi$ and $n_\xi$. A Legendre transform gives then the Hamiltonian~\cite{dempster2014,groszkowski2018} 
\begin{equation}
	\label{eq:Htot}
	H = H_{\theta\phi} + H_\xi + H_{\delta \rm JJ} + H_{\rm int}  + O(\delta C^2, \delta C_J^2).
\end{equation}
with 
\begin{eqnarray}
	H_{\delta \rm JJ} &= - 2E_{C}^{\theta} \delta C_J \; n_\theta n_\phi + E_J \delta E_J \sin(\theta) \sin(\phi + \phi_{\rm ext}/2) \\
	H_{\rm int} &= - 2E_{C}^{\theta} \delta C \, n_\theta n_\xi + E_L \delta E_L\,  \phi\xi.
\end{eqnarray}
The Hamiltonian $H_{\delta \rm JJ}$, depending on $\delta C_J$ and $\delta E_J$, originates from asymmetries between the two JJs. Since the $\xi$ mode does not bias the JJs, junction disorder only affects the capacity matrix of $H_{\theta\phi}$ and distorts the potential $V(\theta,\phi)$. Very strong disorder in the Josephson energies (of the order of $\delta E_J \sim 1$) could completely eliminate the two potential ridges for $\theta=0,\pi$ and hence completely destroy the protection of the qubit. However, the typical experimental disorder in $C_J$ is below $10\%$, while Josephson energies are known to vary from device to device by up to $20\%$ and disorder within the same chip is expected to be significantly smaller \cite{groszkowski2018}. Therefore, even for a conservative estimate of experimental junction disorder, the detrimental effects of $H_{\delta \rm JJ}$ are negligible \cite{groszkowski2018}.\\ 

The situation is radically different for $H_{\rm int}$. There, disorder in the capacitance $C$ and in the inductive energy $E_L$ introduces spurious couplings with the harmonic mode $\xi$, which can have important consequences on the coherence of the whole system. To study their effects, one can resort to standard methods \cite{blais2004,zhu2013}, since the coupling between a qubit and a harmonic oscillator is a typical scenario of circuit QED. In particular, it is convenient to express the whole Hamiltonian as
\begin{equation}
	H = \sum_l \epsilon_l |l\rangle\langle l| + \Omega_\xi a^\dagger a + \sum_{l,l'} (g_{ll'} |l\rangle\langle l'| a + {\rm h.c.}).
\end{equation}
Here, states $|l\rangle$ are the eigenstates of $H_{\theta\phi}+ H_{\delta \rm JJ}$ with energy $\epsilon_l$. Within this notation, the two computational states of the qubit are the ones with $l=0,1$ and the qubit frequency reads $\omega_q = \epsilon_0-\epsilon_1$. Moreover, $a$ ($a^\dagger$) is the creation (annihilation) operator of the harmonic $\xi$ mode, whose frequency is given by $\Omega_\xi  =\sqrt{8E_LE_C^\xi}$. The third term accounts for transitions between the states $|l\rangle$ and $|l'\rangle$ via emission/absorption of excitations of the $\xi$ mode. The strength of the coupling that mediates these transitions is given by $g_{ll'} = g_{ll'}^\phi + i g_{ll'}^\theta$, with \cite{groszkowski2018}
\begin{eqnarray}
	g_{ll'}^\phi &\propto \delta E_L \langle l |\phi |l'\rangle\\
	g_{ll'}^\theta &\propto \delta C \langle l |n_\theta |l'\rangle.
\end{eqnarray}
In the dispersive regime, when the detunings $\Delta_{ll'} = \epsilon_l-\epsilon_{l'}-\Omega_\xi$ are much larger than the couplings $|\Delta_{ll'}| \gg |g_{ll'}|$, the general approach outlined in Ref.~\cite{zhu2013} allows to express the Hamiltonian as 
\begin{equation}
	\label{eq:Hdisp}
	H^{\rm disp} = \sum_l (\epsilon_l +\Lambda_l) |l\rangle \langle l| + \Omega_\xi a^\dagger a + \sum_l \chi_l |l\rangle \langle l| a^\dagger a.
\end{equation}
In this regime, the effects of the interaction between the qubit modes and $\xi$ are captured by the Lamb shift $\Lambda_l$ and the ac Stark shift $\chi_l$, whose expressions read
\begin{eqnarray}
	\Lambda_l = \sum_{l'} \frac{|g_{ll'}|^2}{\Delta_{ll'}}
\end{eqnarray}
and
\begin{eqnarray}
	\chi_l = \sum_{l'} |g_{ll'}|^2  \left(\frac{1}{\Delta_{ll'}} - \frac{1}{\Delta_{l'l}}\right).
\end{eqnarray}
The Lamb shift modifies the qubit frequency $\omega_q$ by a fixed amount $\Lambda_{01} = \Lambda_1-\Lambda_0$ and it is therefore substantially harmless. On the contrary, the frequency shift induced by the ac Stark term, given by $(\chi_1-\chi_0) a^\dagger a $, is proportional to the number of excitations featured by the harmonic mode $\xi$. As a result, (thermal) fluctuations of such a number directly lead to pure dephasing of the qubit. Circuital disorder-induced pure dephasing plays a major role in determining the coherence properties of the $0-\pi$ circuit \cite{groszkowski2018}.\\

One peculiarity of this noise source is its dependence on the ratio $r = E_{C}^{\phi}/E_L$, which is large in the deep regime [see \Eref{eq:deep}]. In particular, there is a (typically high) threshold value $r^{\rm th}$ below which an increase in $r$ can be unexpectedly detrimental for the qubit coherence \cite{groszkowski2018}. For $r>r^{\rm th}$, on the contrary, a further increase of $r$ always leads to an expected reduction of pure dephasing. Such an intriguing behavior, which has to be taken into account for every practical realizations of $0-\pi$ circuits, stems from the existence of two competing phenomena. On one hand, a reduction of $E_L$ (which increases $r$) directly corresponds to a reduction of the harmonic mode frequency $\Omega_\xi = \sqrt{8E_LE_C^\xi}$. This leads to an increase in its thermal population and, thus, in the detrimental fluctuations of the number operator $a^\dagger a$ in \Eref{eq:Hdisp}. On the other hand, if $r$ is large enough, the (approximate) $0-\pi$ symmetry of the circuit starts to strongly and rapidly suppress the qubit's ac Stark coefficient $\chi_{01}=\chi_1-\chi_0$, eventually improving the overall dephasing time \cite{groszkowski2018}.\\

To better put the noise-protection of the $0-\pi$ qubit into perspective, we find it instructive to conclude this discussion by reporting some numerical estimates of noise-related quantities, obtained (theoretically) for reasonable choices of operating parameters. In particular, following Ref.~\cite{groszkowski2018}, we consider
\begin{equation}
	\label{eq:deep_numerical}
	\left\{
	\begin{array}{*{20}{l}}
		E_L/E_J = 4\,\cdot 10^{-3}\\
		E_{C}^{\theta}/E_J \simeq 4\,\cdot 10^{-3} \\
		E_{C}^{\phi}/E_J = 2
	\end{array}\right.
\end{equation}
with $E_J/h = 10$ GHz and a temperature of $T=15$ mK. The frequency of the harmonic mode is then given by $\Omega_\xi/(2\pi)= 113$ MHz and its thermal population is $\langle a^\dagger a\rangle = 2.29$. In this regime, they have found that noise induced by realistic circuital disorder is generally the main source of dephasing, exceeding other possible sources by more than one order of magnitude for almost all the values of the external flux. This is also due to the fact that the parameters in \Eref{eq:deep_numerical} correspond almost exactly to the condition $r\sim r^{\rm th} \sim 500$, i.e. to the scenario when disorder-induced dephasing is maximized. Only a further reduction of $E_L$, therefore, can lead to a significant improvement of the overall coherence. While a more detailed  discussion can be found in Ref.~\cite{groszkowski2018}, this brief analysis already shows how the parameter constraints associated with the deep regime can be stringent and quite challenging to be met in experiments. Indeed, as we detailed in \Sref{sec:0pi_exp}, the first experimental realizations of these systems often operate in a less extreme ``soft'' regime, which can still enjoy partial protection against the noise and constitute important proof-of-concept of the whole design. 
\subsection{Quantum gates on protected qubits}
\label{sec:gates}
As we have discussed above, one of the key ingredients for the intrinsic noise protection of the $0-\pi$ qubit is the exponentially small spatial overlaps of its logical wavefunctions. While this protects the qubit from noisy local perturbations, it also makes it complicate to actively manipulate the quantum state of the qubit. One simple possibility, exploited in some of the recent experiments to characterize a single qubit~\cite{kalashnikov2020}, is to temporarily drive it away from the protected regime and then quickly perform rotations in the Bloch sphere. This approach, however, is prone to decoherence and it is therefore not valid for actual quantum computation, as it jeopardizes the advantages of working with noise-protected qubits. Fortunately, a few more refined strategies have been proposed, which allow for the execution of specific gates with full protection and/or high fidelity {\cite{brooks2013,diPaolo2019,Klots2021}}.

\begin{figure}
	\centering
	\includegraphics[width=0.7\linewidth]{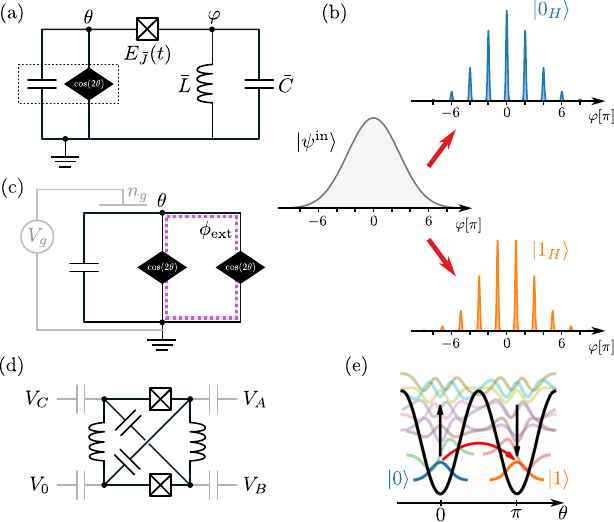}
	\caption{(a) A protected phase gate can be executed by inductively coupling the protected qubit, described here as two superconducting islands that allow for co-tunneling of Cooper pairs between them, with an external LC oscillator, via a tunable JJ (b) Encoding of the qubit state into the \red{Hilbert} space of the harmonic oscillator. According to \Eref{eq:grid}, the harmonic oscillator, initially in its ground state (gray \red{wavefunction}), evolves either to the even (blue) or odd (orange) grid state. (c) Circuit used for implementing the protocols outlined in Ref.~\cite{Klots2021}. Two $\pi$-periodic elements in parallel defines a SQUID-like loop (highlighted in purple), threaded by the external flux $\phi_{\rm ext}$. This, together with the offset charge $n_g$ (in gray), allows for a protected manipulation of the qubit. (d) Qubit control scheme based on capacitive coupling of every node with external (microwave) voltage sources. Capacitances to the ground (usually negligible) are not shown. (e) 1D potential proportional to $\cos(2\theta)$ and sketch of the eigenfunctions $H_{\rm eff}^{\rm 1D}$. The black and red arrows show how to effectively realize a transition from $|0\rangle$ to $|1\rangle$ via the excited states. Adapted from Ref.~\cite{diPaolo2019}.}
	\label{fig:Gates}
\end{figure}

\subsubsection{Fully protected phase gate}

Let us start by reviewing the protected control scheme proposed in Ref.~\cite{brooks2013}. In principle, it can be applied to all $\pi$-periodic qubits described by the effective one-dimensional Hamiltonian $H^{\rm 1D}_{\rm eff}$ in \Eref{eq:H1Deff}. The protocol implements a protected phase gate, which acts on the computational basis by leaving $|0\rangle $ untouched and adding an extra phase to $|1\rangle \to -i |1\rangle$. It thus corresponds to a $\pi/2$ rotation around the $Z$ axis. The general idea is to use a tunable JJ to inductively couple the protected qubit with an external LC circuit with a very large impedance. This allows to temporarily embed the two-dimensional encoded qubit into the infinite-dimensional Hilbert space of the harmonic oscillator, whose dynamics is then used to actually perform the gate. Interestingly, the proposed protocol effectively implements a version of a continuous-variable quantum error-correcting code \cite{gottesman2001}, ensuring exponential suppression of several potential sources of errors throughout the whole execution of the gate. In this respect, a key role is played by the exploitation of so-called grid states of the harmonic oscillators, detailed below and shown in \Fref{fig:Gates}(b).
In what follows, we highlight some important aspects of the protocol, referring the interested readers to the original publication Ref.~\cite{brooks2013} and to Ref.~\cite{gottesman2001} for more detailed discussions.\\

 The qubit is considered as a black box containing a $\cos(2\theta)$ element, described by the effective Hamiltonian $H_{\rm eff}^{\rm 1D}$. Indeed, it only enters the description of the gate protocol as a particular two-terminal circuital element with branch flux $\theta$. As shown in \Fref{fig:Gates}(a), a tunable JJ connects the qubit in series with an LC circuit, the latter being characterized by the flux variable $\varphi$, inductance $\bar L$ and capacitance $\bar C$. The system can be thus described by the Hamiltonian
\begin{equation}
	H(t) = 4 E_{\bar C} n_\varphi^2 + \frac{1}{2} E_{\bar L}\varphi^2 - E_{\bar J}(t) \cos(\varphi - \theta).
\end{equation} 
with $E_{\bar C} = e^2/(2\bar C) $ and $E_{\bar L} = \bar L^{-1} \Phi_0^2 (2\pi)^{-2} $. Assuming the qubit to be operated in the deep regime ($E_2 \gg E_C^\theta$), depending on whether it is in state $|0\rangle$ or $|1\rangle$, the variable $\theta$ is pinned either at $0$ or at $\pi$. One can thus rewrite a qubit state-dependent Hamiltonian just for the $\varphi$ mode
\begin{equation}
	H_{0,1}(t) = 4 E_{\bar C} n_\varphi^2 + \frac{1}{2} E_{\bar L}\varphi^2 \mp E_{\bar J}(t) \cos(\varphi).
\end{equation} 
To perform the gate, at $t=0$, the tunable Josephson coupling ramps on smoothly from zero to a finite value $E_{J_0}$, remains constant for a time $t_0 \sim \hbar (\pi E_{\bar L})^{-1}$, and then ramps off smoothly. \blue{The execution time of the gate is thus proportional to the inductance $\bar L$ of the external LC oscillator. We note that a reasonable value of $\bar L \sim 1\, \mu H$, see \Sref{sec:superinductor}, would correspond to a $t_0$ in the realm of nanoseconds, which is comparable with the execution time of single-qubit gates in standard transmon-based quantum processors \cite{krantz2019}.}

In the beginning, the harmonic oscillator is in its ground state $|\psi^{\rm in}\rangle$. In the limit $E_{\bar C} \gg E_{\bar{L}}$, the Gaussian wavefunction is very broad in $\varphi$ space and the expectation value of the $\cos(\varphi)$ term is thus expected to be exponentially suppressed, i.e. 
\begin{equation}
	\langle \cos \varphi \rangle = e^{-\langle \varphi^2\rangle /2} = \exp\left(
	-\sqrt{\frac{E_{\bar C}}{2E_{\bar L}}}
	\right).
\end{equation}
This ensures that, when the coupling is turned on, the energy of the whole system is basically insensitive to the qubit state or, equivalently, that the $0-\pi$ symmetry of the qubit is (almost) preserved in the low-energy sector by $H_{0,1}(t)$. As $E_{\bar J}$ smoothly ramps up to $E_{J_0} \gg E_{\bar C}$, the Gaussian ground state of the oscillator adiabatically evolves to a ``grid state'', i.e. a superposition of narrow functions peaked around the minima of the cosine potential, still governed by a broad Gaussian envelope [see \Fref{fig:Gates}(b)]. Importantly, the positions of these minima depend on the qubit state. In particular, assuming the initial state of the qubit to be a generic superposition $\alpha|0\rangle + \beta|1\rangle$, by turning on $E_{\bar J}(t)$, the state of the whole system evolves as
\begin{equation}
	\label{eq:grid}
	(\alpha|0\rangle + \beta|1\rangle) \otimes |\psi^{\rm in}\rangle \to \alpha |0\rangle \otimes |0_H\rangle + \beta |1\rangle \otimes |1_H\rangle, 
\end{equation}
where the oscillator grid states $|0_H\rangle$ and $|1_H\rangle$ features peaks for $\varphi$ close to even and odd multiple of $\pi$, respectively. The time evolution of such an entangled state is then controlled by the Hamiltonian $H_{0,1}(t)$ with a constant $E_{\bar J}(t) = E_{J_0}$. Each peak of the grid states is stabilized by the strongly confining cosine potential ($E_{J_0}\gg E_{\bar C}$), but it takes a time-dependent phase $\exp[-it E_{\bar L}\varphi^2/(2\hbar)]$ due to the harmonic potential. After a time $t_0 \sim \hbar (\pi E_{\bar L})^{-1}$, each peak centered around $\varphi \sim n \pi$ accumulates the phase ($m\in\mathbb{Z}$)
\begin{equation}
	\left\{\begin{array}{ll}
		\exp [-i 2\pi m^2] = 1\qquad& {\rm for\, even }\; n=2m \\
		\exp [-i \pi (2m+1)^2/2] =-i & n=2m+1 \\
	\end{array}\right..
\end{equation}
The state of the whole system is then given by
\begin{equation}
	\alpha |0\rangle \otimes |0_H\rangle - i \beta |1\rangle \otimes |1_H\rangle. 
\end{equation}
The execution of the gate is completed by ramping down $E_{\bar J}(t)$ to zero. The state of the harmonic oscillator then evolves to $|j_H\rangle \to |\psi_j^{\rm fin} \rangle$, with $j=0,1$. If $|\psi_0^{\rm fin}\rangle = | \psi_1^{\rm fin}\rangle = |\psi^{\rm fin}\rangle$ (even if $|\psi^{\rm fin}\rangle\neq |\psi^{\rm in}\rangle$), then the system is in a product state and a perfect phase gate has been implemented: 
\begin{equation}
	(\alpha|0\rangle + \beta|1\rangle) \otimes |\psi^{\rm in}\rangle \to (\alpha|0\rangle - i \beta|1\rangle) \otimes |\psi^{\rm fin}\rangle.
\end{equation}

In Ref.~\cite{brooks2013}, the authors showed that a perfect phase gate can be indeed realized, provided that the following conditions are met. First of all, as already mentioned before, it is required to operate in the regime $E_{\bar C}/E_{\bar L} \gg 1$ in order not to break the (almost) perfect degeneracy of the qubit states. Moreover, it is important to properly choose the characteristic time $\tau_J$ for $E_{\bar J}(t)$ to ramp on (and off). Indeed, it must be slow enough to prevent diabatic transitions, which excite the oscillator within each cosine well, while being fast enough to prevent the wavefunction to be squeezed just to a few local minima close to $\varphi \sim 0$. By choosing a time comparable with the period of the oscillator, i.e. for $\tau_J \sim \sqrt{\bar L \bar C}$, both diabatic and squeezing errors are exponentially suppressed with the large ratio $E_{\bar C}/E_{\bar L}$. Also, the effects of small over and under rotations, resulting from the qubit remaining coupled with the oscillator for a too long or short time, do generally not compromise the gate performance. Specifically, the associated gate error is exponentially suppressed with $\sqrt{E_{\bar C} / E_{\bar L}}$, provided that the relative error $\epsilon$ on the uptime of the coupling does not exceed $|\epsilon|<2\pi (E_{\bar C}/E_{\bar L})^{-3/4}$. The protected phase gate is also robust for a sufficiently small finite temperature.

\subsubsection{Towards a universal set of robust gates}
Interestingly, the protocol outlined before can be readily generalized to implement a controlled phase gate between two qubits~\cite{brooks2013}. In order
to obtain a universal set of protected gates, however, single-qubit (non-Clifford) rotations around a different axis are also required \cite{krantz2019,nielsen_book}. A recent proposal by Klote and Ioffe \cite{Klots2021} shows how a relatively small modification of the setup considered by Brooks {et al.} \cite{brooks2013} could lead to universal
quantum computation by allowing for the robust implementation of two additional gates. In particular, they showed how to realize a fully protected $\pi/2$
rotation around the $X$ axis and of a partially-protected but high-fidelity non-Clifford gate, consisting of an arbitrary rotation around the $X$ axis.
The key ingredient that is required for the realization of these two gates {consists in} the possibility to tune the parameter $E_2$ of the effective Hamiltonian
$H_{\rm eff}^{\rm 1D}$. In complete analogy with standard JJs, such a tunability can be achieved by considering two $\pi$-periodic elements
in parallel, which define a dc SQUID-like (superconducting quantum interference device) loop \cite{barone_book} pierced by an external flux $\phi_{\rm ext}$ as sketched in \Fref{fig:Gates}(c). The Josephson energy $E_2$ is thus the largest for $\phi_{\rm ext} = 0$ (mod $2\pi$) while, assuming the two $\pi$-periodic elements to be identical, it can be reduced to zero for $\phi_{\rm ext} = \pi$ (mod $2\pi$).\\ 

To properly describe how the two above-mentioned gates operate, it is important to note that the eigenstates $|+\rangle$ and $|-\rangle$ of the qubit's
$\sigma^x$ Pauli operator have a definite parity of the number of Cooper pairs $n_\theta$ on the floating superconducting island.
Since the ideal deep operational regime of the $\pi$-periodic qubit (i.e. $E_C^\theta\ll E_2$ in $H_{\rm eff }^{\rm 1D}$) is associated with a strong delocalization in charge
space, the wavefunction of the $|+\rangle$ state is thus non-zero for several adjacent even values of the discrete variable $n_\theta$. By contrast, the $-\rangle$
wavefunction is non-zero for several odd values of $n_\theta$. This situation bears strong analogies with the grid states of the harmonic oscillator described
before and, therefore, it allows for a similar (dual) protected manipulation. More precisely, if the Josephson energy $E_2$ is suddenly switched off by
setting $\phi_{\rm ext} = \pi$, the time evolution of the states is exclusively controlled by the capacitive term of the Hamiltonian $H_{\rm eff}^{\rm
	1D}$ via the operator $U(t)=\exp(-i2E_C^\theta t n_\theta^2/\hbar)$. After a time $t_0 \sim \hbar \pi (4E_C^\theta)^{-1}$, the $|+\rangle$ state is brought back to
$U(t_0)|+\rangle = |+\rangle$ while the $|-\rangle$ state picks and extra phase $U(t_0)|-\rangle = e^{-i \pi/2}|-\rangle  =-i |-\rangle$. By suddenly switching
off $E_2$ for a time $t_0$ it is thus possible to implement a $X_{\pi/2}$ gate, featuring analogous exponential protection against timing errors and external
perturbations as the phase gate described before \cite{Klots2021,brooks2013}. \blue{For realistic values of $E_C^\theta$ of the order of several GHz, the gate execution time is expected to be around hundreds of picoseconds.} \\

The non-Clifford rotation is based on a different, holonomic approach \cite{Zanardi1999}. In general, the latter relies on the adiabatic manipulation of
two or more parameters of a qubit that features a degenerate computational manifold. Interestingly, the corresponding unitary operation on the qubit's
state does not originate from dynamical phases (since the computational states are always degenerate) but rather from different (geometric) Berry phases
accumulated during the adiabatic evolution of the system along loops in the parameter space. Such an approach has been proposed as a convenient way to
implement (possibly non-Clifford) gates in several topologically protected platforms, which guarantees the existence of a robust degenerate computational
space \cite{Zanardi1999,Karzig2016,Zhang2018,Zhang2020,vanHeck2012,Calzona2020,Groenendijk2019}. In the specific case under examination \cite{Klots2021},
the tunable parameters that allow for the implementation of holonomic gates are the external flux $\phi_{\rm ext}$ and the offset charge $n_g$, controlled
by the external gate shown in gray in \Fref{fig:Gates}(c), which modifies the capacitive term in $H_{\rm eff}^{\rm 1D}$ as $2E_C^\theta n_\theta^2 \to 2E_C^\theta
(n_\theta-n_g)^2$. These two external knobs are adiabatically controlled to follow a rectangular path in parameter space, ramping up $\phi_{\rm ext} =
0 \to 2\pi$ for $n_g = 0.5$ and then bring it back to $\phi_{\rm ext} = 0$ for a \textit{different} value of the offset charge, $n_g = -0.5$. Along such
a two-dimensional loop, the two states $|+\rangle$ and $|-\rangle$ always remain degenerate but pick a different Berry phase, whose magnitude depends on
the asymmetry between the two $\cos(2\theta)$ elements \cite{Klots2021}. Such an asymmetry can be thus engineered to realize arbitrary non-Clifford rotations.
In the two portions of the loop with $\phi_{\rm ext} \sim \pi$, however, the Josephson energy becomes comparable to (or even smaller than) the charging
energy, $E_2(\phi_{\rm ext})\leq E_C^\theta$, and the qubit ceases to be fully protected. Indeed, the adiabatic evolution of the logical wavefunctions towards this intermediate regime squeezes them in the charge space, thus increasing their sensitivity to fluctuations of $n_g$. As a result, the overall protection
against charge noise-induced dephasing is only limited to the first order. Nevertheless, the non-Clifford holonomic gate is still fully protected against
flux noise \blue{and unwanted diabatic effects are suppressed for gate duration longer than $\tau_{\rm ad} \gtrsim 15 ns$ \cite{Klots2021}. The protocol outlined in this Section has thus} the potential to be implemented with high fidelity, allowing to form a set of (almost) protected universal gates for $\pi$-periodic
qubits \cite{Klots2021}.  

\subsubsection{High-fidelity partially-protected alternatives} 
The actual implementation of the above-mentioned schemes, although extremely promising, comes with several experimental challenges, such as the realization
of very large inductances and/or SQUID-like devices based on $\pi$-periodic elements. Moreover, for the specific $0-\pi$ circuit considered in this Section,
designing an inductive coupling that exclusively addresses the $\theta$ mode might be particularly challenging. Indeed, as highlighted in Ref.~\cite{diPaolo2019},
a connection to a single node flux, say $\phi_A$ in \Fref{fig:0pi}(a), necessarily involves extra unwanted couplings with the $\phi$ or $\xi$ mode [see
Equations \eref{eq:thetaNM}, \eref{eq:xi}, and \eref{eq:phiMN}], potentially spoiling the protection. At the same time, connecting more than one circuit
node to the LC oscillator would shunt some circuital elements, modifying the overall topology of the $0-\pi$ circuit and thus its properties. The development
of gate schemes that allow to fully harness the robustness of protected qubits in realistic settings is thus an important ongoing research area.\\ 

We conclude this Section by briefly describing an alternative protocol that can specifically address the $\theta$ mode of the $0-\pi$
qubit. Despite the lack of full protection, it can potentially lead to implementing fast and high-fidelity single-qubit gates \cite{diPaolo2019}. In contrast
with the previous schemes, for this protocol each node of the $0-\pi$ circuit is \textit{capacitively} coupled to external microwave voltage sources [see
\Fref{fig:Gates}(d)]. The capacitive nature of the couplings preserves the two-island topology of the circuit and does not violate the $0-\pi$ symmetry,
as it only involves charge operators. The drawback is that the exponentially small overlap of the logical qubit states makes conventional control (and
read-out) strategies highly inefficient in the deep regime. As shown in \Fref{fig:Gates}(e), a possible way to circumvent this issue is to consider a gate
scheme that temporarily populates excited states, since the latter feature larger overlaps and can thus mediate a transition from the two logical states.
Since this strategy requires leaving the qubit computational space only for very short times, it can still deliver high-fidelity operations despite a temporary
loss of protection and the need for precise timing of the very short gate temporal duration\blue{, around ten times the inverse plasma frequency of the qubit} \cite{diPaolo2019}. Whether or not the main ingredients of
this protocol can be used to design more involved schemes that allow for multi-qubit (and eventually universal) operations remains an open question.

\section{Experimental realizations}
\label{sec:experiments}
In this Section, we report on some experimental realizations of multi-mode superconducting qubits that have been achieved in the last few years. These have been mostly obtained exploiting conventional superconducting technology, with Al-based junction arrays and Nb islands. Important recent technological advances in fabrication techniques and materials optimization have indeed allowed to obtain proof-of-concept devices of multi-mode protected qubits. 
\subsection{Protected superconducting rhombi chains}
\label{sec:rombi}
\begin{figure}
	\centering
	\includegraphics[width=0.6\linewidth]{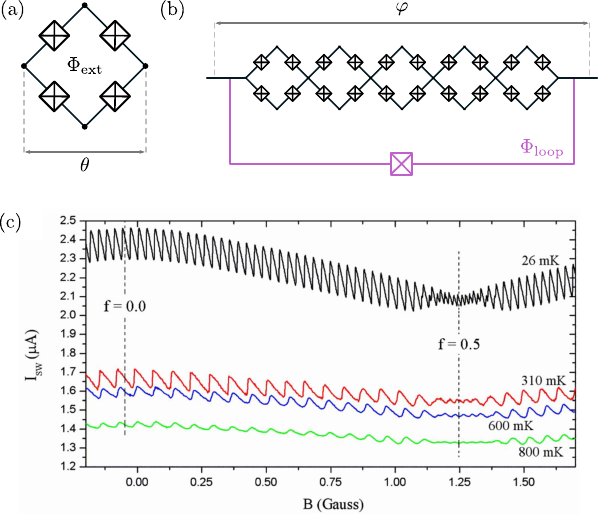}
	\caption{(a) A single rhombus, consisting of four identical Josephson elements that defines a loop, threaded by the external flux $\Phi_{{\rm ext}}$. (b) Example of a rhombi chain consisting of $5$ elements. The phase difference across the whole chain is $\varphi$. To probe the properties of the chain, the latter is shunted by a large JJ (in purple). (c) Experimental plot of the switching current as a function of the external magnetic field, at different temperatures, for a chain containing $8$ rhombi. At full frustration $f=0.5$ (and low temperature) the sawtooth-like modulation of the switching current features a $h/(4e)$ periodicity in the flux threading the larger loop $\Phi_{\rm loop}$. Adapted from Ref.~\cite{pop2008}.}
	\label{fig:rhombi}
\end{figure}

One of the first implementation of protected qubits based on superconducting circuits was proposed in 2002~\cite{ioffe2002, doucot2002, doucot2004}. It relies on a linear chain of superconducting elements, whose single building block is a rhombus made of four JJs, frustrated by an external magnetic flux $\Phi_{{\rm ext}}$, as schematically depicted in \Fref{fig:rhombi}(a).
Every single element constitutes a faulty qubit that at full frustration, i.e. when the transverse magnetic flux is at $\Phi_0 /2$, suppresses coherent transport of conventional $2e$ Cooper pair. This is due to destructive interference of opposite paths along a single rhombus, which resembles the so-called Aharonov-Bohm cages effect~\cite{vidal1998, vidal2001}. It has been shown~\cite{ioffe2002, doucot2002, protopopov2004, protopopov2006} that, although usual $2e$ Cooper pairs result localized, correlated transport along the chain still survives in the form of pair of Cooper pairs with $4e$ charge. This effect originates from the specific geometry of the circuit and it is at the heart of the suppression of conventional single Cooper pairs tunneling, promoting co-tunneling events. It is worth to note that since the capacitance of a single JJ is typically larger than the ground capacitance of islands connecting the different rhombi, the relevant parameters for the rhombi chain are determined by the one of single JJ.\\

In the ideal case, assuming equal JJs, the important energy scales are given by the junction Josephson energy $E_J$ and charging energy $E_C$. Considering a single rhombus, and neglecting the charging energy ($E_J\gg E_C$), the potential profile can be written as the sum of each JJ contributions $\sum_{p=1}^4 E_J[1-\cos(\phi_p)]$, where the phases $\phi_p$ are constrained by the flux across the single rhombus $\sum_{p=1}^4 \phi_p=2\pi\, \Phi_{\rm ext}/\Phi_0 = \phi_{\rm ext}$. Upon minimization, the classic ground state energy $F$ can be written as \cite{doucot2002} 
\begin{equation}
	 - \left|\cos\left(\frac{\theta}{2} + \frac{\phi_{\rm ext}}{4} \right)\right| - \left|\cos\left(\frac{\theta}{2} -\frac{\phi_{\rm ext}}{4}\right)\right|
\end{equation}
where $\theta$ represents a gauge invariant phase (see \Fref{fig:rhombi}(a)). In general, this energy is $2\pi$-periodic both in the phase $\theta$ and in the reduced external magnetic flux $\phi_{\rm ext}$. At half flux $\Phi_{\rm ext}/\Phi_0=1/2$ (or, equivalently, $\phi_{\rm ext}=\pi$), however, the frustration reaches its maximum value and the dependence on $\theta$ becomes $\pi$-periodic. Therefore, at full frustration, we expect two degenerate states to appear, physically describing two oppositely circulating persistent currents. Consistently with the physical picture discussed in \Sref{sec:1deff}, the dominance of co-tunneling of Cooper pairs leads to an effective Hamiltonian of the form of $H^{\rm 1D}_{\rm eff}$ in \Eref{eq:H1Deff}.\\

{A single rhombus, unfortunately, cannot be used as a protected qubit since the $\pi$-periodicity results from the fine tuning of the external magnetic flux and the perfect symmetry between the JJs. Flux noise and fabrication-related disorder are thus extremely detrimental for the performances of a single rohmbus as a qubit. Remarkably, this important limitation can be overcome by considering an array, or chain, of rhombi elements, making the overall system more robust against flux noise and less fragile to any departure from fully symmetric junctions \cite{doucot2012, doucot2002, protopopov2006}. In a rhombi chain, as the one sketched in \Fref{fig:rhombi}(b), the diagonal phase $\theta$ along every single rhombus is linked, 
	and the sum of each $n$-th single diagonal phase $\theta_n$ gives the phase difference between the two ends of the chain $\sum_n \theta_n = \varphi$. At full frustration, i.e. at half magnetic flux per single rhombus, the whole chain behaves as a single effective element featuring $\pi$ periodicity in $\varphi$. It can then be used as a protected qubit, whose sensitivity to flux fluctuations is strongly suppressed with the number of unit cells used in the chain~\cite{doucot2012, doucot2002, protopopov2006}. However, it should be noticed that the critical current of a $n$-elements rhombi chain is $n$ times smaller in amplitude than the one of a single rhombus.}
When charging effects are no more negligible, i.e. when $E_J\gg E_C$ is not satisfied, quantum fluctuations start to play a role. These fluctuations result in quantum phase slip events that, in turn, induce coupling between different states, leading to the formation of macroscopic quantum states extended over the whole chain. This superposition of states lifts the high degeneracy of the classical states~\cite{doucot2002, protopopov2004, protopopov2006}. The presence of quantum phase slips can strongly modify the physical behavior of the chain. Therefore, the ratio $E_J/E_C$ represents a critical parameter for the rhombi chain: one wishes to consider $E_J/E_C\gg 1$ to minimize quantum fluctuations. Indeed, as we saw in \Sref{sec:1deff}, the protected regime of a $\pi$-periodic qubit, described by the effective Hamiltonian $H_{\rm eff}^{\rm 1D}$, guarantees protection against charge noise-induced dephasing only when the effective Josephson energy is much larger than the charging energy. On the other hand, a too large ratio would require a very high degree of accuracy in reaching the full frustration point with the external flux~\cite{doucot2012, protopopov2004, protopopov2006}.\\

In Refs.~\cite{gladchenko2009, pop2008} small size (up to $n=12$) rhombi chains were realized, with a proof-of-concept demonstration of the expected protection, and halved periodicity of the potential, at full frustration. The frustration inside the rhombi chain was
controlled by an external perpendicular magnetic field. In these experiments, the rhombi chain has been fabricated with standard e-beam lithography and shadow evaporation techniques~\cite{krantz2019, barone_book} and consists of optimized Al/AlO$_x$/Al tunnel junctions. The rhombi chain has been introduced in a larger loop containing an additional shunted JJ [see the purple elements in \Fref{fig:rhombi}(b)]. This allowed measuring the current-phase-relation of the chain, by monitoring the critical current, more precisely the switching current, of the parallel circuit. Since the critical current of the larger shunt JJ is much larger than the rhombi chain one, the measured phase dependence directly reflects the current-phase-relation of the chain, allowing for a direct probe of the effective potential periodicity. At the optimal frustration point, the expected behaviour is $E_2\cos(2\varphi)$. 
Small deviations from this regime would introduce additional $E_1\cos(\varphi)$ (conventional) component, that splits the degeneracy of the two minima in the $\cos(2\varphi)$ potential.
These additional features induced by first harmonics have been verified by tuning the external flux away from full frustration $ f = \Phi_{{\rm ext}}/\Phi_0\neq 0.5$~\cite{pop2008}.
Indeed, as reported in \Fref{fig:rhombi}(c) the current-phase relation shows the
characteristic sawtooth-like variation with a periodicity in the flux threading the large loop $\Phi_{\rm loop}$
corresponding to the ordinary superconducting flux quantum $h/2e$
when the rhombi chain is away from full frustration. It turns to half the
flux quantum $h/4e$ at maximum frustration. Moreover, it has been shown~\cite{pop2008} that, by varying $E_J/E_C$, at a lower value of this ratio the appearance of quantum phase slips results in a significant reduction and rounding of the current-phase relation in the
non-frustrated region and complete suppression of the supercurrent at maximal frustration.\\

In Ref.~\cite{bell2014} a modified setup, proposed in Ref.~\cite{doucot2012}, was realized. The improved geometry consists of a common central gate-tunable island shared by two adjacent rhombi, and it becomes very insensitive to the offset charges on all islands except for the central island shared by both rhombi. 
Using a microwave transmission line and a read-out circuit, two-tone spectroscopy, with magnetic flux close to the full frustration regime, was performed to monitor first excitation transitions. From these measurements energy relaxation and hence $T_1$ decay time were reported, with the device tuned to the optimal working point, and compared with performance obtained away from full frustration, in the conventional unprotected regime. The comparison showed that the symmetry protection suppresses the decay rate by almost two
orders of magnitude, reaching values of $T_1 \sim 70\ \mu$s ~\cite{bell2014}. In contrast, a relatively low value of decoherence time was found, with $T_2$ determined by Ramsey interferometer and spin-echo measurements to be $T_2^*\sim 0.45\ \mu$s and $T_{2e}=0.8\ \mu$s. 

\subsection{Large kinetic inductance with superinductors}
\label{sec:superinductor}
As discussed above, a crucial requirement in the development of protected quantum information architectures is to find elements able to efficiently suppress offset charge fluctuations. However, this is not a trivial task with conventional electromagnetic circuits. Indeed, the ratio $Z/R_Q$ between the circuit impedance and the quantum of resistance plays a key role in determining the fluctuation amplitudes of both charges and fluxes. Unfortunately, as previously menioned in \Sref{sec:fluxonium}, the impedence of conventional geometrical capacitance and inductance is typically limited \red{by the vacuum impedance $Z_{{\rm vac}}=\sqrt{\mu_0/\epsilon_0}\sim 377\ \Omega$, which is much smaller than $R_Q$}. This shows that conventional circuit elements easily suffer from large quantum charge fluctuations, whilst being less sensitive to flux ones. \red{Different approaches to largely increase $Z$ are thus required. In this respect, a possible strategy is to exploit the large kinetic inductance of a series of Josephson junctions, a successful approach that we are going to review in the following. Interrestingly enough, however, it has been recently shown that an innovative design of geometric circuital elements, which makes use of mutual inductance of concentric loops, can also lead to impedences exceeding $Z_{{\rm vac}}$ by two orders of magnitude \cite{peruzzo2020,peruzzo2021}.}

In a first attempt to achieve high impedance, the integration of resistive elements and long JJs~\cite{kuzmin1991,lotkhov2003} has been proposed. However, these additional Ohmic components cannot help against charge offset (associated with 1/f noise) and may lead to other dissipation channels, eventually limiting the quantum coherence of devices~\cite{lotkhov2003, masluk2012}.
We note that the kinetic inductance of a single JJ $L_k=(\frac{\Phi_{0}}{2\pi})^{2}/E_{J}$ scales inversely with the Josephson energy $E_J$ and can be thus increased by reducing the in-plane junction dimensions. However, reducing the single junction dimension unavoidably leads to large charging energy $E_C$ and to an exponential growth of the phase-slip rate, causing decoherence.

To bypass the limitations imposed by geometrical constraints, new elements called superinductors have been introduced~\cite{manucharyan2009, manucharyan2012}. These should simultaneously satisfy the following requirements: perfect DC conduction, extremely low dissipation, and very high impedance at relevant frequency ($\sim 1-10$GHz). Superinductors have been recently experimentally realized~\cite{manucharyan2009, manucharyan2012, somoroff2021, masluk2012, pop2012, bell2012, niepce2019, kuzmin2019, gruenhaupt2019, kamenov2020}, exploiting different geometries and platforms, and successfully integrated with superconducting qubits architectures. We now briefly describe their main implementations.\\

\begin{figure}
	\centering
	\includegraphics{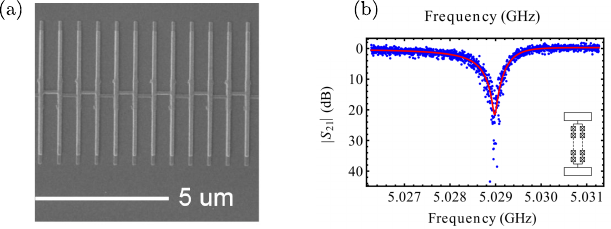}
	\caption{(a) \blue{Scanning electron micrograph image of a section of a Josephson junction array.} (b) Microwave transmission for a $160$-junction array (shown in the inset). The solid line corresponds to a quality factor of $56000$. Both panels are adapted from Ref.~\cite{masluk2012}.}
	\label{fig:ch4si}
	
\end{figure}

The first realizations~\cite{manucharyan2009, manucharyan2012, masluk2012, bell2012} have exploited the large inductance of a series array of JJs. The superinductances are formed by an array of closely spaced JJs (see \Fref{fig:ch4si}(a)) on a SiO$_2$ or sapphire substrate, where the Al-based junctions are fabricated by electron beam lithography and multi-angle shadow mask techniques~\cite{krantz2019, manucharyan2012, barone_book}. The width of the connecting wires between junctions is minimized in order to reduce parasitic capacitances to ground, which ultimately would lower the self-resonant frequency of the superinductance~\cite{hutter2011}. In this configuration, all islands are connected to the rest of the circuit by at least one large junction so that quasistatic offset charges on all islands are screened. The large capacitances of the array junctions prevent phase slips within the array, and for a sufficiently large number (from tens to hundreds) of single junctions the impedance $Z$ can achieve values  greater than $R_Q$. Notice that, in general, a JJ array possesses many degrees of freedom and should be described in a multi-mode scheme. However, in the operating regime of interest, only a relevant harmonic (with quadratic potential) component can be considered.\\ 

The first report~\cite{manucharyan2009} relied on $43$ Al-Ox-Al JJs in series and were used to implement the so-called fluxonium qubit, reviewed in \Sref{sec:fluxonium}. In subsequent experiments~\cite{masluk2012} improved performance where obtained, by increasing of an order of magnitude the $E_J/E_C \sim 100$ ratio and thus suppressing additional decoherence channels due to coherent quantum phase-slip processes.
Low-temperature characterizations of these superinductors were performed by coupling them to a lumped LC resonator capacitively coupled to a transmission line, monitoring their microwave response by means of standard techniques. A typical measurement result is reported in \Fref{fig:ch4si}(b) for a superinductor with $160$ single junction, showing an internal quality factor of the resonator $~56000$, corresponding to a superinductance loss of better than 18 ppm. In that experiment~\cite{masluk2012}, the authors reported superinductances in the range of $100-300$ nH, self-resonant frequencies above $10$ GHz, and phase slip rates below $1$ mHz. The resulting inductance can be extracted from the lowest transition resonance corresponding to $|0\rangle \to |1\rangle $ process via $L = \omega_{0\to 1}^2 C$, reaching values of order $\sim 0.3~\mu$H. Another early realization of superinductors exploiting a combination of multiple JJs has been put forward~\cite{bell2012} in a ladder geometry with the junction frustrated by a magnetic flux. Each unit cell (from 6 to 24 in the experiment) of the ladder, coupled via a large inductance element, forms an asymmetric dc-SQUID comprised of a junction with small $E_J$ in one arm and three junctions with large $E_J$ in the other one. This geometry, when tuned at full frustration with flux at $\Phi_0/2$ across the whole junction, allowed for the maximum fluctuations of the phase
across the ladder and the minimal coupling to the flux noise.\\
In both cases, the reported microwave impedance exceeded the quantum of resistance $R_Q$ by an order of magnitude. In passing, we mention that in Ref.~\cite{pechenezhskiy2020} the JJ array forming the superinductor has been detached from the substrate. By eliminating substrate contributions, stray capacitances have been largely reduced, reaching values of $L=2.5\mu$H and ratios of order $E_L/E_{CL} \sim 0.01$.\\

\begin{figure}
	\centering
	\includegraphics{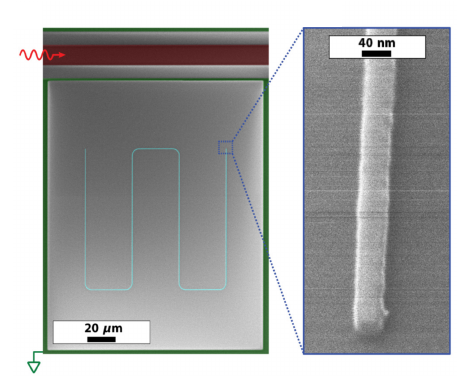}
	\caption{
		False-color SEM micrograph of a nanowire resonator (highlighted in cyan) coupled to a microwave feed line (in red). The NbN ground planes are shown in green, while the exposed Si substrate (where NbN has been etched away) is in gray. A magnified picture of the nanowire is also shown. Adapted from Ref.~\cite{niepce2019}. 
	}
\label{fig:nw}
\end{figure}

Superinductors made of JJ arrays with a large number of unit cells, in addition to their non-trivial fabrication challenges, inherit some limitations due to their intrinsic geometry, in view of possible scalability. An interesting alternative in this respect is constituted by disordered superconducting nanowire~\cite{niepce2019, burnett2017}, whose dimensions can be exploited to limit stray capacitance. However, nanowire geometries are themselves more susceptible to parasitic dissipation mechanisms, i.e. caused by the finite density of localized impurities (often modeled as two-level systems)~\cite{faoro2006, gao2008}, that may limit their applications.
Nevertheless, nanowire-based superinductors have been recently demonstrated using NbN disordered superconductors~\cite{niepce2019}, material that also possesses higher critical fields compared to conventional Al. The investigated devices, shown in \Fref{fig:nw} (a), have $40$~nm width and $680$~$\mu$m length, whose dimensions ensure a large inductance while exponentially suppressing unwanted phase slips~\cite{peltonen2013}.
The value of the kinetic inductance has been extracted from transport measurements of critical temperature $T_c$ and normal resistance $R_n$ via\cite{barone_book}
\begin{equation}
	L=\frac{\hbar R_n}{\pi \Delta}
\end{equation}
with the superconducting gap $\Delta=2.08 k_B T_c$ for NbN superconductors\cite{niepce2019}. There, they obtained $L\sim 2\ {\rm mH}\cdot {\rm m}$, which combined with an estimated capacitance $C\sim 44\ {\rm pF}\cdot{\rm m}$ leads to a device impedance $Z\sim 6.8\ k\Omega$ slightly exceeding $R_Q$. These findings have been also confirmed with microwave spectroscopy, reporting resonator quality factors compatible with the ones of JJ array superinductors, and identifying finite two-level systems density as the predominant source of dissipation. The latter can be further lowered with fabrication improvements such as different substrate choices.\\

Very recently, superinductors have been proposed and fabricated with so-called granular Aluminum (gr-Al)~\cite{gruenhaupt2019, kamenov2020}. This emerging technology can be easily integrated in JJ array geometry and can lead to sizeable improvement in device performance. Indeed, gr-Al simply self-assembles when depositing aluminum in a controlled oxygen atmosphere. Tuning growth parameters, superconducting films with resistivity from $10$ to $10^4\ \mu \Omega \cdot {\rm cm}$ and critical temperature up to $\sim 3$K can be obtained. The wide tunability of the normal state resistance allows to achieve a broad range of high kinetic inductance, retaining high-quality factors up to $10^5$. This new material represents a very promising platform since it would benefit from the easy integration with present Al fabrication technology and from the use of disordered superconductors with higher critical fields, even retaining very low dissipation channels and parasitic contributions.\\

In summary, superinductors are becoming a new key element in superconducting circuit architectures. Indeed, they have the potential to reduce the charge noise sensitivity of Josephson qubits, enable implementation of fault tolerant qubits, and provide sufficient isolation for the electrical current standards. State-of-the-art superinductors, based on substrate-free platform~\cite{pechenezhskiy2020} or gr-Al JJ~\cite{kamenov2020},have still a large potential for improvements and they offer interesting perspective also for wider technological applications such as ultrasensitive single photon detection~\cite{kamenov2020, goltsman2001, kerman2006}.

\subsection{Fluxon parity protected circuit}
We now discuss another approach that allows for the realization of a protected superconducting qubit, known as bifluxon, featuring simultaneous suppression of energy relaxation and flux-noise-induced dephasing. Its working principle is complementary to the one exploited in each element of the rhombi chains introduced in \Sref{sec:rombi}. Indeed, while the latter are based on the Aharonov-Bohm effect, which leads to the suppression of the tunneling of single Cooper pairs, the bifluxon exploits its \textit{dual} effect, the Aharonov-Casher effect~\cite{casher1984}. It predicts that the wavefunction of a moving neutral particle with finite magnetic moment acquires a quantum mechanical phase proportional to the encircled charge. Interestingly, this has been observed also in superconducting junction arrays with $E_J > E_C$~\cite{pop2012, bell2016}, where charges are delocalized and the dynamics can be described in terms of phase slips, i.e. tunneling of fluxons. The idea behind the realization of the bifluxon qubit is thus to design a superconducting circuit where, for a specific value of an external gate voltage controlling the charge of a superconducting island, destructive interference associated with the Aharonov-Casher phase leads to a suppression of single fluxon tunneling. In complete analogy with $\pi$-periodic circuits, the resulting (approximate) conservation of the fluxon parity number can be exploited to encode quantum information in a protected way.\\ 

\begin{figure}
	\centering
	\includegraphics[width=0.7\linewidth]{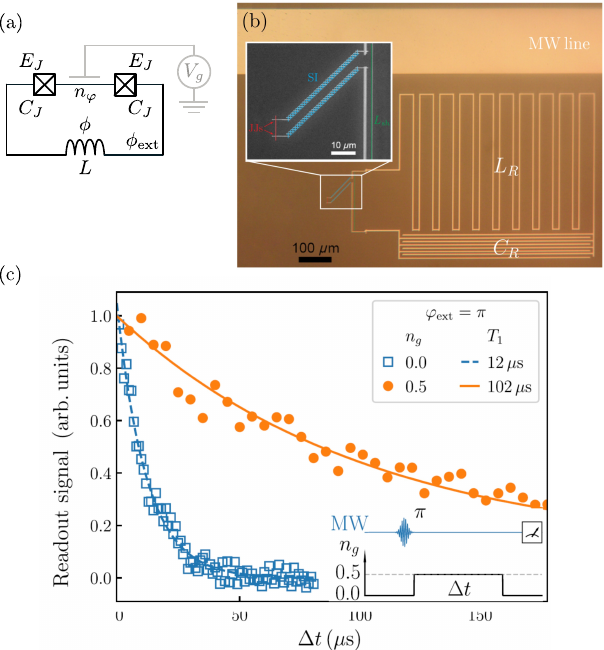}
	\caption{(a) Simplified circuit scheme of the bifluxon qubit. (b) Optical image of the bifluxon qubit, read-out resonator and microwave transmission line. In the inset, a SEM image shows the CPB (in red false color) and the superinductor (in blue), which is actually split into two segments in this particular experimental realization. (c) Measurement of the bifluxon energy relaxation in the protected state (orange) and unprotected state (blue). All panels adapted from Ref.~\cite{kalashnikov2020}.}
	\label{fig:biflux}
\end{figure}

In Ref.~\cite{bell2016}, direct signatures of Aharonov-Casher interference have been measured by inspecting the microwave response of a Cooper pair box (CPB), consisting of two identical JJs separated by a nanoscale superconducting island, shunted by a large superinductor, as sketched in \Fref{fig:biflux}(a). A large kinetic inductance is indeed a key ingredient, as it allows to enter the regime in which the Josephson energy of the junctions $E_J$ dominates over the inductive energy $E_L = (\frac{\Phi_0}{2\pi})^2/L \ll E_J$. In particular, in Ref.~\cite{bell2016}, $L$ exceeds the Josephson inductance by two orders of magnitudes. In this scenario, since one can safely neglect the phase drop on the CPB, the phase difference across the superinductor $\phi$ is completely controlled by the external reduced flux $\phi_{\rm ext}$ that threads the superconducting loop. The energy $\epsilon_m$ of states with a number $m$ of fluxons in the superconducting loop has therefore a parabolic dependence on the external flux, i.e. $\epsilon_m \sim \frac{1}{2} E_L (2\pi m - \phi_{\rm ext})^2$. In general, the crossings between different parabolas are split by phase slips processes on the CPB (i.e. tunneling of fluxons in and out the superconducting loop), which hybridize states with different values of $m$. The suppression of single fluxon tunneling, due to destructive Aharonov-Casher interference, should therefore manifest itself as the disappearance of splittings between parabolas that differ by only one fluxon. This is precisely the experimental signature observed in Ref.~\cite{bell2016}, where two-tone microwave spectroscopy was used to measure the transition frequencies between the ground state and the first two excited states, as a function of the external flux $\phi_{\rm ext}$ and the offset charge $n_g$ of the CPB. The latter is controlled via a gate capacitively coupled to the superconductive island, as sketched in gray in \Fref{fig:biflux}(a). While splittings between the first and the second excited states are always observed, regardless of the value of $n_g$, hybridization between states differing only by one fluxon is strongly suppressed for half-integer values of $n_g$, i.e. when a single electron charge $e$ is present on the superconducting island.\\

The direct observation of Aharonov-Casher interference in Ref.~\cite{bell2016} paved the way for the first realization of a protected qubit based on the conservation of the fluxon number parity, hence dubbed bifluxon~\cite{kalashnikov2020}. The setup is basically identical to the one described before, i.e. a CPB shunted by a large superinductor. A simplified circuit scheme is provided in \Fref{fig:biflux}(a) and an optical image of the bifluxon qubit, together with a read-out resonator and a microwave transmission line, is shown in \Fref{fig:biflux}(b). The CPB is described by the quantized charge of the superconducting island $n_\varphi$ and its conjugated compact flux variable $\varphi$, defined on $(0,2\pi)$. By contrast, the superinductor is characterized by continuous conjugated variables, namely the phase difference $\phi$, introduced before, and the charge $n_\phi$. The whole circuit can be thus described by an effective 2D Hamiltonian
\begin{eqnarray}
	H_{\rm bf} &= 4E_C^\phi n_\phi^2 + 4E_C^\varphi (n_\varphi-n_g)^2 - 2E_J \cos\left(
	\frac{\phi}{2}\right) \cos(\varphi)\nonumber \\& + \frac{E_L}{2} (\phi-\phi_{ext})^2,
\end{eqnarray}
where $E_C^\phi$ and $E_C^\varphi$ are the effective charging energies of the superinductor and the CPB, respectively. Their precise expressions, resulting from the interplay of the self-capacitance of the JJs with stray capacitances to the ground, can be found in Ref.~\cite{kalashnikov2020} together with a detailed quantization of the circuit that takes into account also the read-out resonator. As mentioned before, the offset charge of the CPB $n_g$ is controlled by the voltage of the external gate shown in gray in \Fref{fig:biflux}(a). It is worth noticing that the structure of the Hamiltonian $H_{\rm bf}$, featuring a non-linear coupling between modes $\phi$ and $\varphi$, bears strong similarities with the two dimensional Hamiltonian $H_{\theta\phi}$ of the (symmetric) $0-\pi$ circuit described in \Sref{sec:multimode}.

Let us focus on the limit of full charge frustration $n_g=0.5$ and $E_L\ll E_J$, when the Aharonov-Casher interference suppresses single fluxon tunneling and thus decouples the two sectors with different parity of the fluxon number. In this case, the qubit can be qualitatively described by a fluxon parity-dependent effective 1D Hamiltonian for the $\phi$ mode \footnote{Such a description is quantitatively correct only in the limit $E_C^\varphi \gg E_J$, where it can be derived analytically. Nevertheless, its qualitative features are valid also in the actual operational regime $E_J \sim E_C^\varphi$, where more quantitative analysis necessarily rely on numerical diagonalization \cite{bell2016,kalashnikov2020}}
\begin{equation}
	H_{\pm}^\phi = 4E_C^\phi n_\phi^2 + \frac{1}{2}E_L (\phi-\phi_{\rm ext})^2 \mp E_J \cos\left(\frac{\phi}{2}\right).
\end{equation}
In particular, the even (odd) parity sector is described by $H_+^\phi$ ($H_-^\phi$), whose potential features several deep minima that pin the $\phi$ mode to values $\phi_{m} = 2\pi m$, with $m$ being the even (odd) integer number of fluxons. For a given value of the external flux $\phi_{\rm ext}$, the ground states of the even and odd sectors can then be used to define a qubit, the bifluxon, which features exponential suppression of energy relaxation processes given the disjoint support of the two logical wavefunctions. Moreover, if the logical wavefunctions delocalize over several adjacent minima, the flux dispersion is flattened and the flux noise-induced dephasing becomes exponentially suppressed. To reach this regime, the energy scale associated with fluxon co-tunneling 
\begin{equation}
	E_{\rm ct} \sim \hbar \sqrt{8 E_J E_C^\phi} \exp(-\pi^2 \sqrt{2E_J/E_C^\phi})
\end{equation}
must exceed the inductive energy $E_{\rm ct}\gg E_L$ \cite{kalashnikov2020}. Unfortunately, since these nice properties are the consequence of a precise choice of the offset charge $n_g=0.5$, fluctuations away from this sweet spot are detrimental to the overall performance of the device. In particular, a large charge dispersion makes the bifluxon inherently sensitive to charge-noise-induced dephasing. In this respect, a single bifluxon behaves as the dual of a single rhombus element, introduced in \Sref{sec:rombi}, that features high sensitivity to fluctuations of the external flux away from full frustration sweet spot. Importantly, this analogy also suggests a possible way to improve the robustness of the bifluxon design against charge noise, that is, combining several bifluxon circuits in a small array, thus realizing the dual of a rhombi chain.\\ 

In Ref.~\cite{kalashnikov2020}, a prototype bifluxon device has been fabricated and characterized by an inductively coupled read-out resonator. Obtained device parameters are $E_J=27.2$GHz, $E_C^\varphi=7.7$GHz, $E_L=0.94$GHz, and $E_{C}^\phi=10$GHz. This parameter regime allows for significant suppression of energy relaxation at full frustration, which has been experimentally demonstrated by measuring $T_1$. Indeed, as shown in \Fref{fig:biflux}(c), the relaxation time $T_1$ for $n_g=0.5$ is almost one order of magnitude larger than for $n_g=0$, providing evidence for the qubit's dipole moment suppression due to Aharonov-Casher destructive interference. It is important to stress that the very presence of high protection against relaxation requires the implementation of a non-standard approach to manipulate the qubit and measure $T_1$. In particular, both the $\pi$ pulse that initially populates the qubit excited state and the final measurement are performed in the non-protected regime, i.e. for $n_g \sim 0$. Full charge frustration, and therefore protection against relaxation, is only implemented in between these two operations. Using this particular protocol, values as large as $T_1\sim 100\mu$s have been observed. A similarly modified protocol, where qubit manipulation is performed in the non-protected regime, has been also implemented to measure the overall coherence times, obtaining $T^*_2\sim 0.2\ \mu$s. The latter is negatively affected by the short dephasing time of the device, which traces back to its (inherently) large charge dispersion but also to a fairly large flux dispersion that is associated with the specific parameters used in the experiment. In this respect, a further increase of $L$ (which could be realized using strongly disordered superconducting nanowire based on granular Aluminum~\cite{niepce2019, kamenov2020}) can be particularly beneficial, allowing to access the regime $E_{\rm ct}\gg E_L$ where flux dispersion is exponentially suppressed.
These results have demonstrated the potential of the bifluxon as a protected qubit. Several further improvements can be made in order to reach longer decoherence times, such as minimizing the asymmetry between the JJs of the CPB during the device fabrication. The above-mentioned development of bifluxon arrays also represents a possible strategy to enhance protection against dephasing, in analogy with the rhombi chains. Finally, improved geometries can be engineered to exploit coherent quantum phase slip dynamics~\cite{astafiev2012} and fluxon pairings, along the line of the recently realized dual of a SQUID dubbed charge-quantum interference device~\cite{graaf2018}. 

\subsection{Realization of $0-\pi$ superconducting circuit}
\label{sec:0pi_exp}
Here we discuss the first experimental realization of the $0-\pi$ circuit~\cite{gyenis2021_exp}, whose working principle has been discussed in \Sref{sec:multimode}. We remind the reader that to obtain simultaneous protection from charge and flux noise, this circuit exploits its intrinsic multi-mode nature.
\begin{figure}
	\centering
	\includegraphics{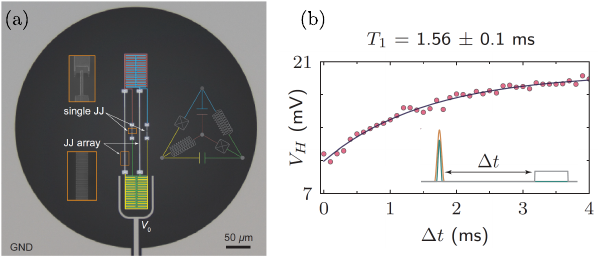}
	\caption{(a) False-color image of the $0-\pi$ qubit. Different colors refer to different nodes (see also the circuits schematics drew on top of the right side of the image). (b) Longitudinal relaxation measurement of the $0-\pi$ qubit. The inset shows the standard pulse scheme. Both panels adapted from Ref.~\cite{gyenis2021_exp}.
	}
	\label{fig:0pi}
\end{figure}
The circuit is composed of a superconducting loop comprising two identical JJs with both large shunting capacitance and shunting inductance, as shown in \Fref{fig:0pi} (a). Analyzing the device geometry, one can identify four nodes, characterized by three relevant degrees of freedom, one of which is a harmonic one. This latter mode can be considered completely decoupled in the ideal case of perfectly symmetric junctions. The resulting Hamiltonian can be described in terms of two variables, called $\theta$ and $\phi$, as in \Eref{eq:Hsym}. As already mentioned, these two modes resemble the dual working principles of capacitively and inductively shunted junctions, respectively. Indeed, $\theta$ is associated with the phase difference between the large capacitors and the JJs (like in a transmon circuit), while $\phi$ represents the phase difference between the superinductor and the JJs (as in a fluxonium device). To reach the protected regime, a huge anisotropy between the two modes is required. In particular, the ideal (deep) regime [see \Eref{eq:deep}] is characterized by strong localization of wavefunctions along $\theta$ direction (often dubbed heavy charge mode) and a wide spread on the opposite $\phi$ profile (often dubbed light flux mode). Therefore, one needs very different charging energies with $E_C^\phi \gg E_C^\theta$, large Josephson energy $E_J\gg E_C^\theta$, and very small inductive energy $E_L\ll E_C^\phi$. We recall that the large Josephson energy is responsible for the formation of a double well profile in the $\theta$ direction, with the resulting wavefunctions localized around $\theta=0$ and $\theta=\pi$~\cite{gyenis2021, gyenis2021_exp}.\\

From an experimental level, the above requirements of energy scales (deep regime) constitute a hard challenge. Indeed, one needs very efficient superinductors to guarantee large kinetic inductance (hence very small $E_L$), precise shaping of capacitors which require a high control degree on unwanted cross capacitance, and JJs as identical as possible to avoid spurious flux contributions. The experiment in Ref.~\cite{gyenis2021_exp} considered a softer regime of operation, characterized by $E_L\sim 0.06 E_J$, $E_C^\theta \sim 0.015 E_J$, and $E_C^\phi\sim 12 E_C^\theta = 0.18 E_J$~\cite{gyenis2021_exp}. These parameters are sufficient to guarantee exponential suppression from charge noise. Moreover, a first-order-insensitive magnetic sweet spot around $\Phi_{{\rm ext}}=0$ has been demonstrated with this finite kinetic energy (which only slightly hybridizes the degenerate doublets on the $\phi$ profile). Also this device has been realized by means of standard lithographic technique and shadow mask deposition. In particular, Nb based capacitors have been placed on a Sapphire substrate at a sufficiently large distance, to limit unwanted cross capacitance contributions, each superinductors is formed by $200$ large JJs, and the two central Al/AlOx/Al JJs have a characteristic Josephson energy of $E_J\sim 6$GHz. Dispersive read-out of qubit performance has been achieved by capacitive coupling with a transmission line resonator with characteristic resonance $\omega_c/(2\pi)\sim 7.2$GHz~\cite{blais2004, blais2021, gyenis2021_exp}. Two-tone spectroscopy as a function of offset charge $n_g$ and external flux $\Phi_{{\rm ext}}$ has been employed to determine the spectrum of the circuit, finding a remarkable agreement with the simple two-mode Hamiltonian $H_{\theta\phi}$ in \Eref{eq:Hsym}.\\ 

{Direct transitions between the computational states are strongly suppressed, because of their disjoint support. The control of the qubit has been therefore achieved by taking advantage of higher-energy states, as the latter have support on both the $\theta=0$ and $\theta=\pi$ valleys [see \Fref{fig:Gates}(e)]. While this is reminiscent of the protocol proposed in Ref~\cite{diPaolo2019} and mentioned in \Sref{sec:gates}, these two schemes are actually based on different strategies. In particular, the experimental implementation relied on Raman processes that virtually populate one ancillary excited state. This has the advantage that only a very small fraction of the population occupies the unprotected intermediate state but, unfortunately, its applicability is limited to the soft regime since, in the deep one, the onset of an approximate selection rule makes Raman-based gates ineffective~\cite{diPaolo2019}.
	This limitation is overcome by the deep regime-ready protocol described in \Sref{sec:gates}, where several excited states are actually populated for a very short time. Nevertheless, for current experimental realizations of the $0-\pi$ qubit, a Raman gate scheme is effective in controlling population transfer between the two computational states, as demonstrated by clear Rabi oscillations reported in Ref.~\cite{gyenis2021_exp}. There, an impressive relaxation decay time $T_1\sim 1.6$ms has been also recorded, as shown in \Fref{fig:0pi}(b), which is larger than state-of-the-art transmon performance \cite{place2021,wang2021}, and comparable to the last achievement in fluxonium device \cite{somoroff2021}. As for the overall coherence times, obtained by Ramsey interferometry and Hahn echo protocol, values of $T_2^* \sim 9\ \mu$s and $T_{2e} \sim 25\ \mu$s have been extracted, demonstrating first order protection from flux noise \cite{gyenis2021_exp}.} This first experimental realization of $0-\pi$ circuit confirms the huge potential of these protected superconducting circuits, opening new perspectives for the implementation of protected multi-qubits gates and more complex operations.

\subsection{Kinetic interference cotunneling element}
We now describe the implementation of a generalized Josephson element~\cite{smith2022}, first proposed in Ref~\cite{smith2020}, called Kinetic Interference co-Tunneling Element (KITE). Also this superconducting circuit is able to enter a protected regime, where the transport is dominated by co-tunneling events of pair of Cooper pairs. Furthermore, by investigating its flux response, it is possible to find a direct link between protected regime and an enhancement of zero-point phase fluctuations. This fact therefore reduces the asymmetry between charge and flux fluctuations discussed in \Sref{sec:fluxonium}, paving the way for artificial systems where the two can act in a symmetric fashion.\\

\begin{figure}
	\centering
	\includegraphics{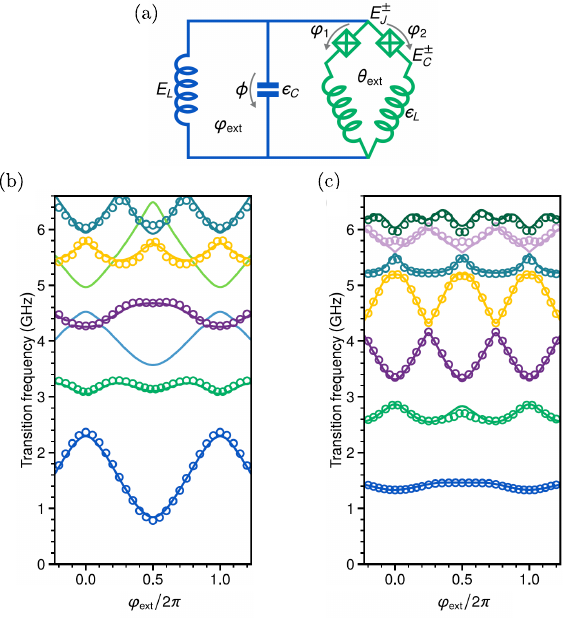}
	\caption{(a) Schematic of the KITE device. (b,c) Two-tone spectroscopy measurements of the transition frequencies from the ground state, for $\theta_{\rm ext} = 0$ (b) and $\theta_{\rm ext} = \pi$, respectively. All panels adapted from Ref.~\cite{smith2022}}
	\label{fig:kite}
\end{figure}

The circuit consists of a superconducting loop where a parallel of two small identical JJs (with Josephson energy $E_J$ and charging energy $E_C$) is placed in series with two superinductances (with associated energy $E_L$), see the green portion of the sketch in \Fref{fig:kite}(a). The parameter regime implemented here is $E_L < E_{CL} \ll E_J \sim E_C$ with $E_L=0.23$GHz, $E_{CL}=2.5$GHz, $E_J=5.9$GHz, and $E_C=6.6$GHz.  
This configuration, in contrast with the rhombus element~\cite{doucot2012, gladchenko2009, bell2014}, does not contain any central superconducting island, and it is thus less sensitive to offset charge noise.
Thanks to the presence of an additional shunting large inductance (see blue region in the figure), this device can pass from a conventional $2\pi$ to a $\pi$ periodicity, depending on the external flux threading the whole circuit. The latter behavior occurs at half flux and it is caused by Aharonov-Bohm destructive interference of single Cooper pair tunneling, exactly as in the case of a rhombus element~\cite{doucot2002, protopopov2004}.
 We notice that in the experimental realization both superinductors have been realized by JJ arrays containing $100$ single large JJ elements by standard double angle Al evaporation technique. The circuit has three degrees of freedom, two associated with the KITE and one with the shunting portion. Since $E_C\sim E_J$ is the dominant energy scale, the system can be effectively described in terms of only two phases by relying on the Born-Oppenheimer approximation. Tuning the external flux from 0 to half flux $\theta_{{\rm ext}}=\pi$, the reduced Hamiltonian can be further simplified with the following one-dimensional effective description
\begin{equation}
	H_\mu= 4 E_C \left(\frac{n}{\mu}\right)^2 + \frac{E_L}{2}(\varphi - \varphi_{{\rm ext}})^2 + (-1)^\mu E_J \cos (\mu \varphi)~,
\end{equation}
where $\varphi_{{\rm ext}}$ is associated to the external flux across the suprconducting loop and the parameter $\mu=1,2$ determine the two possible case, respectively. Note that the $(-1)^\mu$ factor guarantees that the cosine potential is pinned to 1 at $\pm\pi$ for all $\mu=1,2$. Device properties have been measured with an inductively coupled LC read-out resonator. Circuit transition frequencies have been probed by two-tone spectroscopy. It has been demonstrated~\cite{smith2022} the halving of periodicity, from conventional $2\pi$ at zero flux to $\pi$ at half flux.\\

 Correspondingly, when the period is halved the ground state wavefunction delocalization gets magnified. This is confirmed by the suppression of flux dispersion of the first excited state transition, with a ten-time reduction when passing from zero to half flux [see \red{blue circles in }\Fref{fig:kite}(b,c)]. Indeed higher level transitions have a strong flux dependence, consistent with the ground state and first state delocalization over several Josephson potential wells. Notably, this delocalization magnification is also linked to an enhanced amplitude of zero-point-fluctuations of the phase across the KITE, which is estimated as
\begin{equation}
	\varphi_{\mathrm{zpf},\mu} = \left(\frac{2E_{C}}{E_L}\mu^2\right)^{1/4}\;.\label{eq:phizpf} 
\end{equation}
From the experiment a factor $\varphi_{{\rm zpf},2}/\varphi_{{\rm zpf},1}\sim 2.1$ enhancement has been reported. The fact that this value is larger than the expected $\sqrt{2}$ one, has been ascribed to the presence of double phase slip events when the external flux is zero (in the conventional $2\pi$ periodicity regime)~\cite{smith2022}.\\

These results can have important fundamental consequences in finding systems where charge and flux fluctuations are no more asymmetric. Indeed, the expression for zero-point-fluctuations can be rewritten as
\begin{equation}
	\varphi_{\mathrm{zpf},\mu} = \sqrt{8 \pi \alpha_{{\rm ee}} \mu \frac{Z}{Z_\mathrm{vac}}}
\end{equation}
where $\alpha_{{\rm ee}}=1/137$ is the fine structure constant and $Z$, $Z_{{\rm vac}}$ represent the circuit and vacuum impedance, respectively. 
We indeed recall that the ratio between flux and charge fluctuations is of order $8\alpha_{{\rm ee}}$.In addition, in a circuit where electrons can only move in packets of $\mu$ Cooper pairs, one would expect that the ratio of flux to charge fluctuations gains a factor of $\mu$. This, in turn, implies that more complex superconducting circuits where transport is mediated solely by correlated entities of multiplies of Cooper pairs will further lower the ratio between flux and charge fluctuations, paving the way for a novel class of superconducting devices with built-in symmetric noise response.

\section{Conclusions}
 \label{sec:concl}
 
 Thanks to their versatility and high-degree of control, superconducting
 circuits are at the forefront of novel quantum technology applications.
 Among all superconducting based qubits are a promising and powerful
 resource for the ensuing quantum computing era. After the recent
 demonstration of working devices able to perform non-trivial quantum
 computational tasks~\cite{arute2019, preskill2018}, in the so-called
 noise intermediate-scale quantum framework, worldwide theoretical and
 experimental efforts are devoted in finding strategies to completely
 suppress, or at least to mitigate, errors and noise sources. Here we
 focused on novel circuit designs, i.e. multi-mode superconducting
 circuits, that exploit their  topology to achieve noise protection at a
 hardware level. In essence, the larger number of degrees of freedom,
 with respect to conventional single-mode qubits,  is at the root of
 robust encoding of quantum information against environmental
 disturbances. Different proof-of-concept experiments have recently
 demonstrated the huge potential of such multi-mode qubits, achieving
 performance comparable, or even greater than, state-of-the-art
 single-mode superconducting qubits.
 One of the next key steps in building quantum information platforms
 consist in translating noise protection on a larger scale, together with
 the development of modular and versatile hardware and circuit couplers
 that can guarantee efficient information transfer and high-fidelity of
 different computational tasks~\cite{chou2018,campbell2022,schrade2022}. A modular approach can indeed manage complex systems by
 assembling smaller components into larger networks, possibly performing
 operations in noisy resilient sub-parts and transferring information via
 efficient communication channels. \red{An additional application of protected superconducting qubits can be devised by
 integrating them with conventional transmon devices in a heterogeneous quantum processors. The protected qubit in this situation could act as a memory, while the transmon could provide high-fidelity gate operations \cite{maiani2021,ciani2022}.} \red{More generally, an interesting perspective} would be the integration of
 superconducting circuit technology with hybrid
 superconductor/semiconductor based devices~\cite{schrade2022,larsen2020,guo2022}. Here, the large versatility offered by
 semiconductors, both in geometrical control and in the possible gate
 tunability, can indeed further enhance the performance of
 superconducting based qubits in a already mature technological context.

 \section*{Acknowledgments}
A.C. was supported by the W\"{u}rzburg-Dresden Cluster of Excellence ct.qmat, EXC2147, project-id 390858490, and the DFG (SFB 1170). A.C. also thanks the Bavarian Ministry of Economic Affairs, Regional Development and Energy for financial support within the High-Tech Agenda Project ``Bausteine f\"{u}r das Quanten Computing auf Basis topologischer Materialen''.


\end{document}